\documentclass[aip,jcp,twocolumn,reprint]{revtex4-1}

\usepackage{graphics,graphicx,amsfonts,amsmath,amsbsy,amssymb,color}
\usepackage{bm}
\usepackage[a4paper,vmargin={20mm,20mm},hmargin={20mm,10mm}]{geometry}
\usepackage{subfigure}
\usepackage{paralist}
\usepackage{bbold}
\usepackage{mathtools, nccmath}

\def \Schrodinger {{Schr\"{o}dinger }}
\def \Lowdin {{L\"{o}wdin}}

\newcommand{\bra}{\langle}
\newcommand{\ket}{\rangle}
\newcommand{\bs}[1]{\boldsymbol{#1}}

\def \Efn {{E^{\textrm{fn}}}}
\def \Hfn {{\hat{H}^{\textrm{fn}}}}
\def \Hfnp {{\hat{H}^{\mathrm{fn\prime}}}}
\def \Psifn {{\Psi^{\textrm{fn}}}}
\def \PsiT {{\Psi^{\textrm{T}}}}
\def \psiiT {{\psi_i^{\textrm{T}}}}
\def \psijT {{\psi_j^{\textrm{T}}}}
\def \psiiT {{\psi_i^{\textrm{T}}}}
\def \VSFi {{\mathcal{V}_i^{\textrm{sf}}}}
\def \ELi {{E^{\textrm{L}}_i}}

\begin{document}

\title{Fixed and partial-node approximations in Slater determinant space for molecules}
\author{Nick~S.~Blunt}
\email{nicksblunt@gmail.com}
\affiliation{Yusuf Hamied Department of Chemistry, Lensfield Road, Cambridge, CB2 1EW, United Kingdom}
\affiliation{St John's College, St John's Street, Cambridge, CB2 1TP, United Kingdom}

\begin{abstract}
We present a study of fixed and partial-node approximations in Slater determinant basis sets, using full configuration interaction quantum Monte Carlo (FCIQMC) to perform sampling. Walker annihilation in the FCIQMC method allows partial-node simulations to be performed, relaxing the nodal constraint to converge to the FCI solution. This is applied to \emph{ab initio} molecular systems, using symmetry-projected Jastrow mean-field wave functions for complete active space (CAS) problems. Convergence and the sign problem within the partial-node approximation are studied, which is shown to eventually be limited in its use due to the large walker populations required. However the fixed-node approximation results in an accurate and practical method. We apply these approaches to various molecular systems and active spaces, including ferrocene and acenes. This also provides a test of symmetry-projected Jastrow mean-field wave functions in variational Monte Carlo (VMC) for a new set of problems. For trans-polyacetylene molecules and acenes we find that the time to perform a constant number of fixed-node FCIQMC iterations scales as $\mathcal{O}(N^{1.44})$ and $\mathcal{O}(N^{1.75})$ respectively, resulting in an efficient method for CAS-based problems that can be applied accurately to large active spaces.
\end{abstract}

\date{\today}

\maketitle

\section{Introduction}
\label{sec:intro}

Quantum Monte Carlo (QMC) methods provide a powerful approach to solve electronic structure problems, particularly when high accuracy is needed. One of the most widely-used QMC methods is variational Monte Carlo (VMC)\cite{McMillan1965}, in which one guesses the form of the desired wave function, and can sample its variational energy (and other properties) through a Monte Carlo approach such as the Metropolis algorithm\cite{Metropolis1953}. Typically, the wave function contains parameters such that the wave function can be further optimized by a variety of approaches\cite{Sorella2001, Umrigar2005, Umrigar2007, Schwarz2017}. However, the form of this VMC wave function is still somewhat limited, and finding accurate and compact representations of the exact wave function is challenging. Further methods are therefore used to improve this VMC solution. A common method for this task is diffusion Monte Carlo (DMC)\cite{Grimm1971, Anderson1975, Umrigar1993, Foulkes2001}. DMC is a projector QMC approach, which will converge upon the best wave function whose nodal surface equals that of the trial wave function. DMC is a continuum-space method, where sampling is performed in the position-space representation of the wave function.

In contrast, most electronic structure methods are performed within a finite basis set, most commonly Slater determinants formed from molecular orbitals. This truncates the full Hilbert space, and the exact solution within this truncated basis is known as full configuration interaction (FCI). This approach has both its advantages and disadvantages, but its use is extremely common, such that QMC methods working within this representation are valuable.

VMC is less widely performed in determinant space, compared to its continuum-space counterpart. This is partly due to the relative expensive of the method, having both a higher scaling and prefactor. However, it has recently been shown that both the scaling and prefactor can be bought down significantly, resulting in a much more practical method\cite{Wei2018, Sabzevari2018}. This is an important development, especially since many wave functions in determinant-space VMC can be applied to strongly correlated systems with high accuracy and polynomial-scaling computational cost with system size.

Similarly, the fixed-node approximation has been less widely used in determinant-space QMC methods\cite{vanBemmel1994, tenHaaf1995, Becca2017}, compared to DMC. Some of the reasons for this are the same; the former method typically requires accumulating all connected determinants from each walker, and so is expensive. Because of this, it has largely been applied to lattice models. It also recovers less of the remaining error compared to DMC. Nonetheless, it is an important way to improve the accuracy of VMC, where it can be afforded. We note that auxiliary-field quantum Monte Carlo (AFQMC) is another important projector QMC method within finite basis sets\cite{Zhang2003, Motta2018, Malone2020, Shi2021}. AFQMC typically employs the phaseless approximation, which is very different to the fixed-node approximation, and we do not consider it here.

Separately, full configuration interaction quantum Monte Carlo (FCIQMC) has been developed in recent years\cite{Booth2009}. It has similarities to existing projector QMC methods, in particular its evolution of walkers by the imaginary-time \Schrodinger equation. However, the approximation applied is very different to that made in most QMC approaches. Specifically, FCIQMC usually relies on the initiator approximation\cite{Cleland2010, Cleland2011}, which applies a truncation on the Hamiltonian (dynamically changing based on the distribution of walkers), and therefore has more in common with truncated configuration interaction (CI) approaches. Importantly, the FCIQMC method can achieve stable sampling for systems with a sign problem, which is possible through its annihilation procedure\cite{Spencer2012}.

In addition to the initiator approximation, it is also possible to perform the fixed or partial-node approximation in FCIQMC, and there have been two previous studies of this idea. The Fermi polaron was studied by Kolodrubetz and Clark\cite{Kolodrubetz2012}, and the uniform electron gas was studied by Roggero \emph{et al.}\cite{Roggero2013} using a coupled cluster doubles wave function. Applying these ideas to \emph{ab initio} molecular systems is challenging for a few reasons: the cost of accumulating all connected determinants for each walker; the difficulty of obtaining accurate VMC wave functions; and the technical challenge of having access to an efficient implementation of FCIQMC and such VMC wave functions in the same code.

In this article we build upon the recent developments for determinant-space VMC in molecular systems. In particular, due to developments by Neuscamman and Sharma and co-workers, wave functions have been developed that are particularly accurate for strong correlation, and the cost of performing VMC in this way has been reduced\cite{Neuscamman2012, Neuscamman2013, Neuscamman2016_2, Wei2018, Sabzevari2018, Mahajan2019}. This provides an opportunity to investigate projector Monte Carlo as a technique to improve the accuracy of this approach further, and simultaneously provides an opportunity to improve the FCIQMC method by making use of these accurate trial wave functions. This leads us to develop a fixed-node FCIQMC method which gives accurate results as a CASCI solver, and has low polynomial scaling with the active space size. For a series of acenes, we observe an approximate scaling of $\mathcal{O}(N^{1.75})$ with active space size, for a fixed number of FCIQMC iterations.

In Section~\ref{sec:theory} we review the relevant theory, including the fixed-node approximation, and define the VMC wave functions used. We then discuss how FCIQMC differs from its traditional implementation when performing fixed and partial-node approximations. Computational details are defined in Section~\ref{sec:computational_details}, and the systems and active spaces used are defined in Section~\ref{sec:systems}. In the results, Section~\ref{sec:results}, we first study the sign problem with importance sampling and the partial-node approximation applied. We then apply the fixed-node approximation to a range of systems and active spaces, and finally discuss the computational cost and scaling of the method. In Section~\ref{sec:comparison}, we compare node-based approaches to initiator FCIQMC and other related quantum chemical methods.

\section{Theory}
\label{sec:theory}

\subsection{The fixed-node approximation in finite basis sets}
\label{sec:fixed_node}

We begin with an overview of the fixed-node approximation in finite basis sets. In particular, this has commonly been applied to lattice models, but can also be considered for FCI problems in $\emph{ab initio}$ molecular systems.

Most projector Monte Carlo methods can only be performed in the absence of a sign problem. Such sign problems only occur in situations where walkers of both positive and negative sign can be created on a given configuration. One common way to avoid this is by truncating the Hamiltonian, setting to zero any elements $H_{ij} = \bra D_i | \hat{H} | D_j \ket$ which flip a sign relative to those in a trial wave function, which we denote $| \PsiT \ket$. In the determinant basis $\{ | D_i \ket \}$ we denote components of the trial wave function by
\begin{equation}
| \PsiT \ket = \sum_i \psiiT | D_i \ket.
\end{equation}

It is simple to define a sign-problem-free Hamiltonian, $\Hfnp$, parameterized by a real number $\gamma$,
\begin{equation}
H^{\textrm{fn}\prime}_{ij}(\gamma) =
\begin{cases}
    H_{ii},           & \text{for } i = j\\
    H_{ij},           & \text{for } i \ne j, \;\; s_{ij} < 0\\
    - \gamma H_{ij},  & \text{for } i \ne j, \;\; s_{ij} > 0,
\end{cases} 
\end{equation}
where $s_{ij} = \psiiT H_{ij} \psi_j^{\mathrm{T}}$. If $s_{ij} < 0$ then a spawning between $|D_i\ket$ and $|D_j\ket$ will not introduce a sign flip relative to the trial wave function. If $s_{ij} > 0$, then the Hamiltonian element is replaced by $- \gamma H_{ij}$, preventing a sign problem for $\gamma \ge 0$. For $\gamma = -1$ the exact Hamiltonian is recovered.

Using $\Hfnp$, energies obtained from the standard QMC estimator will be non-variational in general. Therefore, it is common to work with a modified Hamiltonian instead, introduced by van Bemmel \emph{et al.}\cite{vanBemmel1994} This is the fixed-node Hamiltonian, $\Hfn$, defined by\cite{vanBemmel1994, Sorella2000, Sorella2002, Becca2017}
\begin{equation}
H_{ij}^{\mathrm{fn}}(\gamma) =
\begin{cases}
    H_{ii} + (1 + \gamma)\mathcal{V}_i^{\mathrm{sf}},  & \text{for } i = j\\
    H_{ij},                                            & \text{for } i \ne j, \;\; s_{ij} < 0\\
    - \gamma H_{ij},                                   & \text{for } i \ne j, \;\; s_{ij} > 0
\end{cases} 
\end{equation}
where $\mathcal{V}_i^{\mathrm{sf}}$ is the \emph{sign-flip potential} at $| D_i \ket$, defined by
\begin{equation}
\mathcal{V}_i^{\mathrm{sf}} = \sum_{j : s_{ij} > 0} {H}_{ij} \frac{\psi_j^{\mathrm{T}}}{\psiiT},
\end{equation}
with a contribution added to the diagonal for each sign-violating connection from determinant $| D_i \ket$.
The lowest eigenstate and eigenvalue of $\Hfn$ define the fixed-node wave function and energy, denoted
\begin{equation}
\Hfn(\gamma) | \Psifn(\gamma) \ket = \Efn (\gamma) | \Psifn(\gamma) \ket.
\end{equation}
The fixed-node energy $\Efn (\gamma)$ can be exactly sampled by the usual mixed estimator available to projector Monte Carlo methods,
\begin{equation}
\Efn(\gamma) = \frac{ \bra \PsiT | \, \hat{H} \, | \Psifn(\gamma) \ket }{ \bra \PsiT | \Psifn(\gamma) \ket  }.
\end{equation}
The equality of this estimator with $\Efn (\gamma)$ follows because $\hat{H} | \PsiT \ket = H^{\mathrm{fn}}(\gamma) | \PsiT \ket $.

A lot can be said about the form of $\Efn(\gamma)$ as $\gamma$ is varied. In particular\cite{tenHaaf1995},
\begin{equation}
E_0 \le \frac{ \bra \Psifn(\gamma) | \, \hat{H} \, | \Psifn(\gamma) \ket }{ \bra \Psifn(\gamma) | \Psifn(\gamma) \ket  } \le \Efn(\gamma) \le E^{\mathrm{T}}
\end{equation}
for any $\gamma \ge -1$, such that the fixed-node approximation is both variational and an improvement to the VMC energy of trial wave function, $E^{\mathrm{T}}$. Lowering $\gamma$ is also guaranteed to improve the energy estimate,
\begin{equation}
\frac{d\Efn(\gamma)}{d\gamma} \ge 0,
\end{equation}
with the exact result obtained at $\gamma=-1$. Lastly, $\Efn(\gamma)$ is a concave function for real $\gamma$,\cite{Beccaria2001} allowing an improved variational estimate to be obtained by a linear extrapolation to $\gamma=-1$ from two values $\Efn(\gamma_1) > E_0$ and $\Efn(\gamma_2) > E_0$.

While $\Hfn$ has many of these desirable properties over $\Hfnp$, it is not necessarily true that $\Hfn$ will give significantly more accurate energies or properties in a given example. It is not clear the extent to which this is true, for the types of \emph{ab initio} problems to be studied here. Therefore, while we primarily focus on $\Hfn$, we will also look at results obtained with $\Hfnp$.

\subsection{Importance sampling}

It is common to apply importance sampling when working with $\Hfn$. This is widely used in most projector QMC methods, but has not been commonly applied in FCIQMC (although it has recently been applied to sign-problem-free systems in FCIQMC\cite{Ghanem2021}). The elements of the importance sampled fixed-node Hamiltonian in the determinant basis are defined by
\begin{equation}
\tilde{H}^{\mathrm{fn}}_{ij} = \psiiT H^{\mathrm{fn}}_{ij} \frac{1}{\psi_j^{\mathrm{T}}}.
\end{equation}
If the components of the exact ground state of $\bs{H}^{\mathrm{fn}}$ are $\psi_i^{\mathrm{fn}}$, then the components for the ground state of $\tilde{\bs{{H}}}^{\mathrm{fn}}$ are $\psiiT \psi_i^{\mathrm{fn}}$, and the energies are identical. Note that importance sampling does not modify the diagonal elements of $\bs{H}^{\mathrm{fn}}$, such that the sign-flip potential is unchanged.

\subsection{Trial wave functions}
\label{sec:trial_wf}

To apply the fixed-node approximation, accurate trial wave functions are needed for which the components $\psiiT = \bra D_i | \PsiT \ket$ can be calculated efficiently.

We use symmetry-projected Jastrow mean-field wave functions, which take the form
\begin{equation}
| \PsiT \ket = \hat{J} \hat{P} | \phi \ket.
\label{eq:trial_wf}
\end{equation}
Here, $| \phi \ket$ is a mean-field wave function, that will be either a generalized Hartree--Fock (GHF) or an antisymmetric geminal power (AGP) wave function, as defined below. $\hat{P}$ is a projector that restores certain symmetries broken in $| \phi \ket$, and $\hat{J}$ is a Jastrow factor.

These wave functions were investigated in detail recently by Mahajan and Sharma in ref~\onlinecite{Mahajan2019}, and here we directly use the same approach. We also use the same VMC approach to sample and optimize these wave functions, described in ref~\onlinecite{Sabzevari2018}. Readers are referred to these articles for an in-depth discussion. Here we briefly define the relevant points as needed.

The GHF wave function is defined by
\begin{equation}
| \phi^{\mathrm{GHF}} \ket = \prod_{i=1}^{N} \Big( \sum_{p\sigma} \theta_i^{p\sigma} \hat{a}^{\dagger}_{p\sigma} \Big) | 0 \ket,
\end{equation}
where $\hat{a}_{p\sigma}^{\dagger}$ creates an electron in the orbital labelled $p$ with spin $\sigma$. The parameters $\theta_i^{p\sigma}$ are optimized as part of the VMC procedure, obtaining the optimal orbitals in the presence of the Jastrow factor. GHF generalizes the orbital definition such that they are no longer separable into spin and spatial parts. This increases the variational freedom of the wave function compared to restricted Hartree--Fock (RHF), at the cost of breaking $\hat{S}_Z$ and $\hat{S}^2$ symmetries.

The AGP wave function\cite{Coleman1965, Casula2003, Sorella2007, Neuscamman2012, Henderson2019} is defined by
\begin{equation}
|\phi^{\mathrm{AGP}}\ket = \Big( \sum_{pq} F_{pq} \hat{a}_{p\uparrow}^{\dagger} \hat{a}_{q\downarrow}^{\dagger} \Big)^{N/2} |0\ket.
\end{equation}
where $\bs{F}$ is the pairing matrix. We choose $\bs{F}$ to be symmetric to enforce singlet spin for the AGP, in contrast to the GHF wave function.

For the results in this article both $\bs{\theta}$ and $\bs{F}$ are allowed to be complex, which increases the variational freedom of both ansatz.

A `density-density' form for the Jastrow is used, defined as
\begin{equation}
\hat{J} = \mathrm{exp} \Big( \sum_{i \ge j} \nu_{ij} \hat{n}_i \hat{n}_j \Big),
\label{eq:jastrow}
\end{equation}
where $i$ and $j$ are spin orbital labels, and $\hat{n}_i$ is the number operator for the spin orbital with label $i$. The Jastrow has $(M+1)M/2$ parameters, where $M$ is the number of spin orbitals.

The projection operator $\hat{P}$ restores the broken $\hat{S}_Z$ and complex conjugation ($\hat{K}$) symmetries (although $\hat{S}_Z$ is only broken in the GHF reference), and so $\hat{P} = \hat{P}_{S_Z} \hat{P}_K$. In practice, $|\PsiT\ket$ only appears through its overlap with a determinant, which automatically enforces $\hat{S}_Z$ symmetry. Following the choice from ref~\onlinecite{Mahajan2019}, we refer to these two wave functions as J-K$S_Z$GHF and J-KAGP.

Consider the action of the Jastrow operator onto a determinant. If the orbitals in the determinant are the same as those in the Jastrow factor, then this is trivial to calculate. However, if the orbitals in the determinant are different to those in the Jastrow factor, then there is no efficient way to exactly calculate this overlap. This has an important impact on the results in this article, because the orbitals used in the FCIQMC simulation must therefore be the same as those chosen for the Jastrow in the VMC simulation. The optimal choice of orbitals for the Jastrow factor is an open question, but a very sensible (and probably near-optimal) choice are localized orbitals. By using localized orbitals, Neuscamman proved that the Jastrow AGP wave function can be made exactly size consistent\cite{Neuscamman2012}. In our investigations, we always find localized orbitals to be optimal over any other systematic construction, as measured by the optimized VMC energy. Therefore, this choice of orbitals is also used in FCIQMC simulations, except where stated otherwise.

These wave functions are often very accurate for capturing strong correlation, particularly in complete active space problems. However they have low accuracy for treating dynamical correlation in large orbital basis sets\cite{fn1}. Therefore in this article we focus on solving CASCI problems. Dynamical correlation could be included through multireference perturbation theory, which can also be performed with VMC\cite{Blunt2020}.

\subsection{FCIQMC treatment of $\Hfn$}

In this study we have used the FCIQMC method to perform fixed and partial-node approximations. The fixed-node Hamiltonian has previously been used primarily in the Green's function Monte Carlo (GFMC) method\cite{Ceperley1979, Trivedi1989, vanBemmel1994, Sorella2000, Becca2017}. FCIQMC has a number of differences that are interesting to study in the context of nodal approximations. In particular, the annihilation procedure in FCIQMC allows stable sampling in the presence of a sign problem, provided a sufficient walker population is used. This will allow us to partially turn off the nodal approximation, to converge to the exact result at $\gamma=-1$. Such an approach has previously been investigated by Kolodrubetz and Clark\cite{Kolodrubetz2012}, where the Fermi Polaron was studied. Roggero \emph{et al.} studied the uniform electron gas using the fixed-node approximation and a coupled cluster doubles wave function\cite{Roggero2013}. Here, we extend these ideas to \emph{ab initio} molecular systems using J-K$S_Z$GHF and J-KAGP wave functions.

In FCIQMC, the wave function is represented by a collection of walkers. The total amplitude of walkers on $|D_i\ket$ is denoted $C_i$, so that the FCIQMC wave function is
\begin{equation}
| \Psi \ket = \sum_i C_i |D_i\ket.
\end{equation}
In the original presentation of FCIQMC, $C_i$ were taken to have integer values, but have since been extended to non-integer values, usually referred to as real coefficients\cite{Petruzielo2012}. We use real walker coefficients in this paper.

These walkers are evolved by spawning and death rules, so as to sample the imaginary-time \Schrodinger equation,
\begin{equation}
C_i(\tau + \Delta\tau) = C_i(\tau) - \sum_j (H_{ij} - S \delta_{ij}) C_j(\tau).
\label{eq:itse}
\end{equation}
where $S$ is a shift that is added to the diagonal of the Hamiltonian, and varied to control the walker population. In the limit of large $\tau$, the walker coefficients will sample the ground state of $\hat{H}$, provided $\Delta\tau$ is sufficiently small. Importantly, the rules are such that the expectation value of walker amplitudes at $\tau+\Delta\tau$ exactly follow Eq.~(\ref{eq:itse}), so that the algorithm is unbiased, despite statistical error in the propagation. The FCIQMC algorithm has been detailed numerous times, and we refer readers to previous studies for a precise definition.\cite{Booth2009, Spencer2012, Petruzielo2012}. Here, we detail differences necessary to treat the fixed-node Hamiltonian, $\Hfn$.

With importance sampling applied, the walkers amplitudes will represent $C_i = \psiiT \psi_i$. We use the following estimator,
\begin{align}
E &= \frac{\bra \PsiT | \hat{H} | \Psi \ket}{\bra \PsiT | \Psi \ket}, \\
  &= \frac{\sum_{ij} \psiiT H_{ij} \psi_j}{\sum_i \psiiT \psi_i}, \\
  &= \frac{\sum_i C_i E^{\mathrm{L}}_i}{\sum_i C_i},
\end{align}
where $\ELi = \sum_j H_{ij} \frac{\psijT}{\psiiT}$ is the local energy of $| \PsiT \ket$ on determinant $|D_i\ket$ (taking real Hamiltonian elements).

A difficulty arises when using the fixed-node Hamiltonian including the sign-flip potential, $\Hfn$. If $\psiiT$ is very small for a given determinant, then the sign-flip potential $\VSFi$ can become arbitrarily large. If the following is satisfied,
\begin{equation}
\Delta\tau (H_{ii} + \VSFi - S) > 1,
\label{eq:death_error}
\end{equation}
then the death step will cause the walker amplitude to change sign, and possibly explode, and the simulation will become unstable. In GFMC, this is avoided by using a continuous time algorithm. In theory it is possible to do the same in FCIQMC, and such approaches have been investigated\cite{Kolodrubetz2012, smart_thesis}. However, this leads to a very different algorithm, and makes annihilation much more difficult to perform. Since we are keen to investigate annihilation and the sign problem, we avoid this approach. Instead, we make an approximation, following the suggestion of Kolodrubetz and Clark\cite{Kolodrubetz2012}, where we replace the death step by
\begin{equation}
1 - \Delta \tau (H_{ii}^{\mathrm{fn}} - S) \rightarrow e^{-\Delta\tau (H_{ii}^{\mathrm{fn}} - S)},
\end{equation}
replacing the linearized projection operator by its exponential form. For an extremely large $\VSFi$, this will have the effect of killing the walker. Importantly, we only apply this approximation in cases where Eq.~(\ref{eq:death_error}) is satisfied. In practice, this is extremely rare, so that this approximation has negligible effect on final estimates, which we have confirmed by testing with and without the approximation applied. We believe that this a better strategy than the alternative of setting the time step $\Delta\tau$ extremely small, which greatly reduces the efficiency of the method.

Another difference to traditional FCIQMC simulations is in the nature of the basis used. Typically, \emph{ab initio} FCIQMC simulations are performed in a canonical or natural orbital basis, where the wave function expansion is dominated by relatively few determinants. However, using a localized orbital basis means that the CI expansion converges very slowly\cite{fn2}. In this basis, the annihilation rate is much lower. This affects the usefulness of various approaches developed in FCIQMC. In particular, the replica trick\cite{Overy2014} is much less effective, such that variational and perturbative\cite{Blunt2018, Blunt2019} estimators are too noisy to obtain accurately. The semi-stochastic approach is also not useful, as it relies on a finding a small deterministic space which dominates the CI expansion\cite{Petruzielo2012}. In some results we will investigate the use of canonical and split-localized orbital basis sets, where the CI expansion converges quickly, in which case these approaches are useful. However note that the chance of Eq.~(\ref{eq:death_error}) being satisfied is much larger in this case, and the density-density Jastrow factor is less accurate.

A potentially expensive step is the calculation of the local energy $E^{\mathrm{L}}_i$ and sign-flip potential $\VSFi$ for each occupied determinant $|D_i\ket$. Indeed, the number of connected determinants scales as $\mathcal{O}(M^4)$ with the number of orbitals, $M$. In FCIQMC we usually try to avoid looping over all connections due to this prohibitive expense. However, this expense can be greatly reduced by using the heat bath approach\cite{Holmes2016_2, Sharma2017, Sabzevari2018},
\begin{equation}
\ELi \approx \sum_j^{\epsilon} H_{ij} \frac{\psijT}{\psiiT},
\label{eq:local_E_HB}
\end{equation}
where contributions are excluded for which $|H_{ij}| < \epsilon$. Considering just double excitations, where $H_{ij}$ takes the form $\bra ab || pq \ket$, only $\mathcal{O}(M^2)$ elements will be non-negligible when using localized orbitals, in the large system limit. A similar argument can be made for single excitations. This approach is discussed in ref~\onlinecite{Sabzevari2018}. The same screening is also applied to $\VSFi$. The cost requirement in FCIQMC is also reduced by storing $\ELi$ and $\VSFi$ for each occupied determinant, avoiding recalculation. Since FCIQMC is time-limited far more than memory-limited, this is a sensible choice. Note that we choose to not apply this screening in the FCIQMC spawning step, although this may be interesting to investigate in future work.

Lastly, we point out that VMC can be used to initialise the FCIQMC walker distribution from a sampling of $|\PsiT\ket$. When applying the importance-sampled Hamiltonian, the trial wave function is represented by $C_i = (\psiiT)^2$, which is the distribution sampled by the VMC algorithm. Similarly, when importance sampling is not in use, one can place a walker of amplitude $1/\psiiT$ on each selected determinant to sample the same wave function.

\section{Computational details}
\label{sec:computational_details}

We use PySCF\cite{pyscf, pyscf_2020} to perform initial RHF, GHF and CASSCF calculations, and to perform orbital localization. Where geometry optimizations were performed, PySCF was used with the geomeTRIC library\cite{geometric}. For the CASSCF calculations, heat bath CI (HCI)\cite{Holmes2016_2, Sharma2017, Smith2017} is used as an approximate solver with the Dice code\cite{Dice}. DMRG benchmarks were generated with BLOCK\cite{Chan2002, Chan2004, Ghosh2008, Sharma2012, Olivares2015}. VMC calculations were performed using the Sharma group code\cite{VMC_GitHub, Sabzevari2018, Mahajan2019}, using AMSGrad\cite{AMSGrad} to optimize VMC wave functions. We developed a new implementation of the FCIQMC algorithm within this VMC code, available on GitHub\cite{VMC_GitHub}, allowing efficient access to trial wave functions and the required overlaps.

When calculating the local energy or sign-flip potential, as in Eq.~(\ref{eq:local_E_HB}), we usually take $\epsilon=10^{-8}$ Ha, except where stated otherwise.

\section{System and active space definitions}
\label{sec:systems}

Here we define the systems and active spaces that are studied in Section~\ref{sec:results}, and define how the CASCI problems are set up.

In each case where CASSCF is performed to obtain the final orbitals used, HCI was used as an approximate solver\cite{Smith2017}. The exception is ferrocene, where an exact solver was used. In Section~\ref{sec:results}, VMC and fixed- and partial-node FCIQMC will then be used to solve the resulting CASCI problems accurately.

Linear hydrogen chains H$_n$ are studied in the STO-6G basis, an ($n$e,$n$o) active space, and an internuclear distance of R=2 $a_0$.

For acenes, we optimized the ground-state geometry using the 6-31G* basis and with CAM-B3LYP\cite{CAM-B3LYP}. We then performed CASSCF with a 6-31G basis. In each case, the active space consists of all valence $\pi$ orbitals, from ($10$e,$10$o) for napthalene to ($34$e,$34$o) for octacene.

For the 9,10-bis(phenylethynyl)anthracene (BPEA) molecule, we take the geometry from ref~\onlinecite{Blunt2019_2}, and perform CASSCF, with a threshold of $\epsilon = 5 \times 10^{-5}$ Ha for the HCI solver, and the cc-pVDZ basis. The active space consists of all valence $\pi$ orbitals, ($30$e,$30$o).

For trans-polyacetylene (TPA), C$_{2n}$H$_{2n+2}$, as for linear acenes, we perform CASSCF using a 6-31G basis and an active space consisting of all valence $\pi$ orbitals, ($2n$e,$2n$o).

For ferrocene, we reproduce the ($10$e,$7$o) and ($18$e,$15$o) active spaces from ref~\onlinecite{Knizia2017}, including the same basis and level of theory to set up each CASCI problem. In particular, restricted open-shell Hartree--Fock (ROHF) is performed with the cc-pVTZ-DK basis and the exact-two-component (x2c) approach. Then the active space is formed using the atomic valence active space (AVAS) method, using a threshold of $0.2$ (see ref~\onlinecite{Knizia2017} for details). For the ($10$e,$7$o) active space the target set of atomic orbitals are the $3d$ orbitals of Fe. For ($18$e,$15$o), the target set is the $3d$ orbitals of Fe and $2p$ orbitals of C atoms. CASSCF is then performed, using an exact solver. The $D_{5h}$ geometry from ref~\onlinecite{Harding2008} is used.

For Fe(II)-Porphyrin (Fe(P)) we take the geometry and active space of Smith \emph{et al.}\cite{Smith2017}. This active space was first investigated by Li Manni \emph{et al.}\cite{Manni2016}, and consists of $20$ C $2p_z$, $4$ N $2p_z$ and 5 Fe $3d$ orbitals, giving a ($32$e,$29$o) active space. CASSCF orbitals were obtained using a cc-pVDZ basis.

We investigated different orbital localization schemes. Intrinsic bond orbitals (IBOs)\cite{Knizia2013} were used for ferrocene, \Lowdin's $\bs{S}^{-1/2}$ procedure was used for H$_n$ chains, and the Foster-Boys approach\cite{FosterBoys} was used for all other systems.

All geometries are given in the Supporting Information.

\section{Results}
\label{sec:results}
\subsection{Importance sampling and the sign problem}
\label{sec:plateaus}

\begin{table*}
\begin{center}
{\footnotesize
\begin{tabular}{@{\extracolsep{4pt}}lccc@{}}
\hline
\hline
Orbitals used & Importance sampling applied? & Walker population at plateau & Annihilated walkers per iteration \\
\hline
Localized        &  No   & $6.8 \times 10^5$ & $400$ \\
Localized        &  Yes  & $5.7 \times 10^5$ & $860$ \\
Split-localized  &  No   & $5.5 \times 10^6$ & $2.8 \times 10^4$ \\
Split-localized  &  Yes  & $7.9 \times 10^7$ & $2.0 \times 10^6$ \\
Canonical        &  No   & $5.1 \times 10^6$ & $9.5 \times 10^4$ \\
\hline
\hline
\end{tabular}
\caption{Plateau heights and annihilation rates for the FCIQMC simulations presented in Figure~\ref{fig:plateau_diff_orbs}, performed on H$_{14}$ in a STO-6G basis with R=2 $a_0$. The annihilation rates are calculated during the plateau region. Values are given to $2$ significant figures.}
\label{tab:plateau_heights}
}
\end{center}
\end{table*}

Before investigating the fixed-node approximation, it is interesting to investigate what effect importance sampling has on the sign problem in FCIQMC. Spencer \emph{et al.} showed that the severity of the sign problem in FCIQMC is related to the difference in eigenvalues between two matrices.\cite{Spencer2012} Define the transition matrix $\bs{T} = -(\bs{H} - S\bs{I})$, and matrices containing only its positive and negative elements, $\bs{T}^+$ and $\bs{T}^-$. Then the sign problem can be understood as related to the difference in the largest eigenvalues between matrices $\bs{T}^+ - \bs{T}^-$ and $\bs{T}^+ + \bs{T}^-$. It is simple to prove that the importance sampling transformation leaves the eigenvalues of \emph{both} matrices unchanged. Therefore, we perhaps should not expect any significant change in the sign problem severity. However, this severity is also determined by the rate of annihilation and other factors, and it is not clear what effect importance sampling may have here.

\begin{figure}
\includegraphics{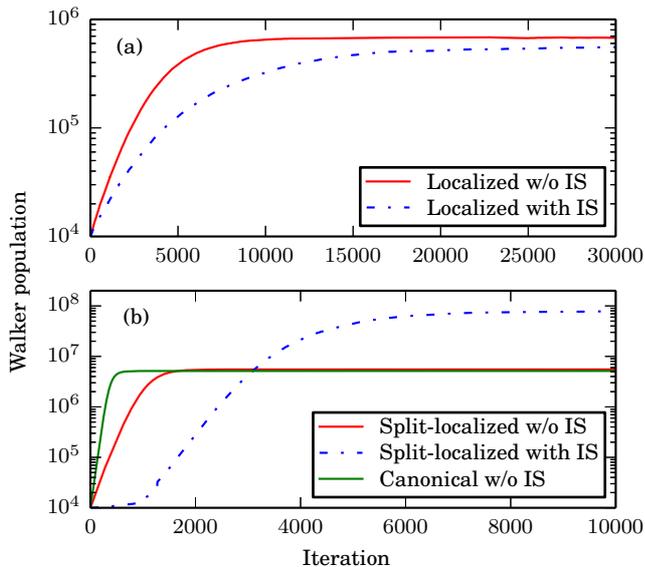}
\caption{Walker population plateaus for H$_{14}$ (R=2 $a_0$, STO-6G) with different orbitals, with and without importance sampling applied. The plateau height is a metric of sign problem severity. (a) Using localized orbitals, (b) using canonical and split-localized orbitals. With localized orbitals importance sampling is seen to slightly reduce the plateau height and the severity of the sign problem. In the case of split-localized orbitals, importance sampling worsens the sign problem severity. Interestingly, the sign problem is less severe with localized orbitals than with canonical orbitals.}
\label{fig:plateau_diff_orbs}
\end{figure}

The sign problem severity in FCIQMC can be assessed by the walker population plateau height. This determines the minimum number of walkers required to achieve stable sampling in the presence of a sign problem. To investigate this we look at H$_{14}$. In Figure~\ref{fig:plateau_diff_orbs}, different molecular orbitals are investigated. This affects the orbitals used in both the Jastrow factor \emph{and} the FCIQMC determinants. Thus it affects the accuracy of the trial wave function and also the basis used. The space size is $\sim 10^7$ determinants. Using localized orbitals, the plateau heights are $5.7 \times 10^5$ and $6.8 \times 10^5$ with and without importance sampling, respectively. In this case, importance sampling slightly reduces the sign problem severity.

To aid with understanding, Table~\ref{tab:plateau_heights} presents annihilation rates for the simulations in Figure~\ref{fig:plateau_diff_orbs}. In the definition used, a walker of weight $+1$ annihilating with a walker of weight $-1$ counts as $2$ annihilated walkers. With localized orbitals in use, it is seen that the annihilation rate is indeed higher when importance sampling is applied, which is consistent with a lower plateau height.

We also investigated using canonical and split-localized orbitals. In the latter, the occupied and virtual orbitals are localized separately. Note that one would usually obtain the natural orbitals before performing this localization, but here for simplicity we perform the localization on the canonical RHF orbitals. It is known that split-localized orbitals can lead to a particularly fast convergence of the FCI expansion, comparable to using natural orbitals\cite{Bytautas2003}. This is in contrast to fully-localized orbitals, where the FCI expansion converges extremely slowly. Since localized orbitals lead to an accurate Jastrow of the form in Eq.~(\ref{eq:jastrow}) but a slow CI convergence, split-localized are an interesting compromise to investigate.

Without importance sampling, split-localized orbitals and canonical orbitals lead to a similar plateau height around $\sim 5 \times 10^6$ walkers. Notably, the sign problem here is more severe than with localized orbitals. One might imagine that the sign problem is particularly severe in the localized orbital basis, as the annihilation rate is much lower, as seen in Table~\ref{tab:plateau_heights}. Our results show that this intuition is incorrect, and that the sign problem can be ameliorated by using localized orbitals. Some related examples of this are well known; in particular the 1D Hubbard model is sign-problem-free in some instances in a basis of local orbitals, but not when working in a basis of Bloch functions. However it is not clear that the sign problem should be less severe in \emph{ab initio} systems, as seen here.

When importance sampling is applied in the split-localized basis, the plateau height increases to $\sim 8 \times 10^7$ walkers. These results together show that importance sampling can either improve or worsen the sign problem in FCIQMC, so that there is no systematic improvement. In Table~\ref{tab:plateau_heights} it is seen that the annihilation rate at the plateau is much higher in this case after applying importance sampling. Therefore the increased plateau height cannot solely be attributed to a differing annihilation rate. More generally we have found that importance sampling has a negative effect on statistics in canonical and split-localized basis sets. This can be understood because the ratios $\psiiT/\psijT$ between connected determinants can differ by several orders of magnitude in this case, leading to erratic spawning behaviour. It is possible that this contributes to the increased sign problem severity that is observed here, although it is challenging to assess this conclusively. We note that a similar importance sampling method has been discussed in the density matrix quantum Monte Carlo (DMQMC) method\cite{Blunt2014}. A recent study has also observed an increased plateau height when importance sampling is applied in DMQMC\cite{Petras2021}, consistent with our results. In summary, importance sampling is recommended when working with localized orbitals, but not with canonical, split-localized or natural orbitals.

\subsection{Partial-node approximations}

\begin{figure}
\includegraphics{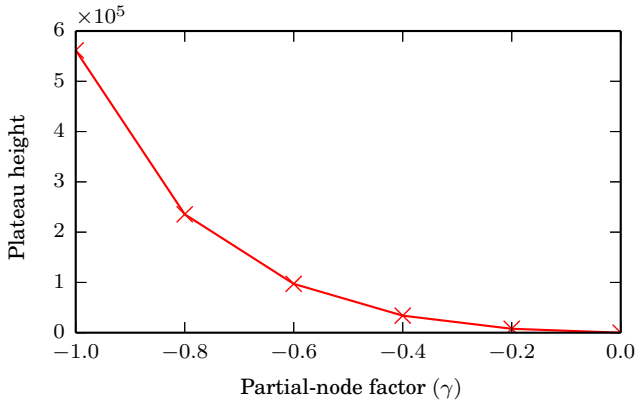}
\caption{Walker population plateaus heights for H$_{14}$ (R=2 $a_0$, STO-6G) with the partial-node approximation applied, using the Hamiltonian $\Hfn(\gamma)$. $\gamma=-1$ is the exact Hamiltonian, while $\gamma=0$ is sign-problem-free.}
\label{fig:plateau_partial_node}
\end{figure}

Next we consider partial-node approximations, where $-1 < \gamma < 0$, leading to a sign problem of increasing severity as $\gamma$ approaches $-1$.

We again begin by investigating walker population plateaus in H$_{14}$, with results shown in Figure~(\ref{fig:plateau_partial_node}). Here, we use the Hamiltonian $\Hfn$, which includes the sign-flip potential. It can be seen that the plateau height decreases quite quickly as $\gamma$ is increased toward $0$, faster than a linear decay. The plateau height at $\gamma=-1$ is $5.4 \times 10^5$ walkers, whereas at $\gamma=-0.8$ the plateau height is around $2.4 \times 10^5$ walkers. As such, there may be cases where one cannot perform fully free propagation, but can achieve near-exact results with a partial node approach.

We have also investigated equivalent results when using the Hamiltonian $\Hfnp$, which does not include sign-flip potential. Here we tend to find that the sign problem is slightly more severe than that found with $\Hfn$, requiring slightly larger walker populations to reach the plateau.

Next we investigate energies within the partial-node approximation both with and without the sign-flip potential. We look at three systems: H$_{14}$ ($14$e,$14$o); anthracene ($14$e,$14$o); and trans-polyacetylene (TPA) C$_{12}$H$_{14}$ ($12$e,$12$o). These are small active spaces, which allow us to investigate $\Efn(\gamma)$ to the exact result at $\gamma=-1$. A J-K$S_Z$GHF trial wave function was used in each case. For the first two systems, (a) and (b), we use localized orbitals, which is the standard procedure. For TPA C$_{12}$H$_{14}$ the results presented use split-localized orbitals.

\begin{figure}
\includegraphics{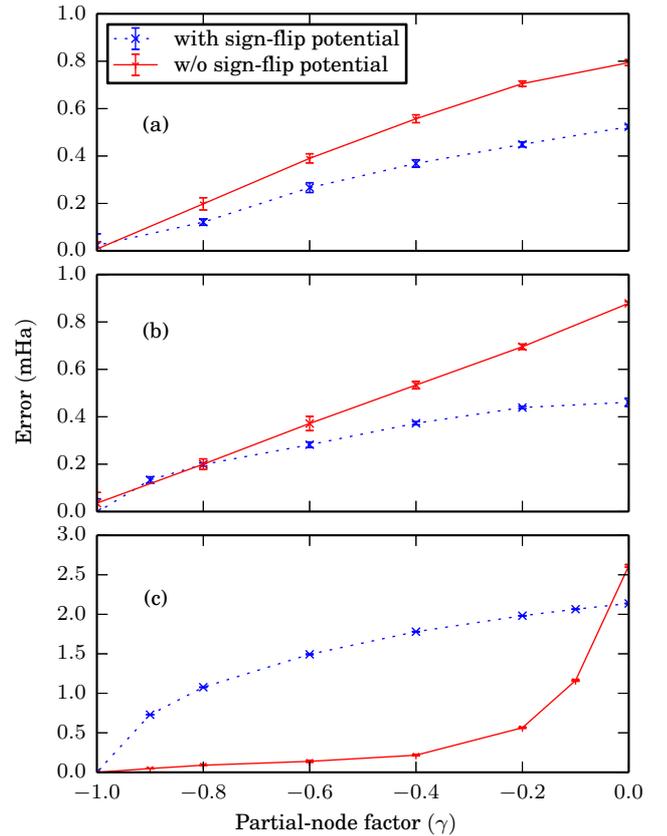}
\caption{The partial-node approximation applied both with and without the sign-flip potential applied. A J-K$S_Z$GHF trial wave function was used. Localized orbitals were used for (a) and (b), while split-localized orbitals were used for (c). The systems are: (a) H$_{14}$ STO-6G (R=2 $a_0$) (14e,14o) (b) Anthracene, full $\pi$-valence active space (14e,14o). (c) Trans-polyacetylene C$_{12}$H$_{14}$, (12e,12o) active space.}
\label{fig:partial_node}
\end{figure}

Results are presented in Figure~(\ref{fig:partial_node}). In cases (a) and (b) the J-K$S_Z$GHF wave function and the fixed-node approximation ($\gamma=0.0$) are seen to be extremely accurate, giving better than $1$ mHa accuracy. In (c) the error is slightly larger, which is a consequence of using split-localized orbitals, but still better than $3$ mHa with $\gamma=0$. Reducing $\gamma$ is found to systematically improve energies in all cases. This is a known result for the Hamiltonian $\Hfn(\gamma)$, as described in Section~\ref{sec:fixed_node}. However, it is also seen to be true here for the Hamiltonian $\Hfnp$. Results for $\Hfnp$ are also all found to be variational.

For systems (a) and (b), where localized orbitals are used, including the sign-flip potential is found to improve energies for almost all values of $\gamma$. However, the improvement is not always dramatic. Interesting results are observed for system (c), where split-localized orbitals are used. While $\Hfn$ gives a better energy than $\Hfnp$ at the sign-problem-free point, $\gamma=0$, energies for the latter Hamiltonian converge much more quickly. It seems that $\Hfn$ can lead to a limited improvement and slow convergence with respect to $\gamma$ when using canonical or split-localized orbitals, while $\Hfnp$ leads to faster convergence in this case. As described in Section~\ref{sec:fixed_node}, $\Efn(\gamma)$ is known to be concave. Nonetheless, overall the results are seen to be extremely accurate, even at $\gamma=0$.

We can also take this opportunity to look at the difference in accuracy in the J-K$S_Z$GHF wave function between orbital basis sets. For C$_{12}$H$_{14}$ (12e,12o) we find errors of 0.239(5) mHa, 3.832(7) mHa and 17.47(2) mHa for localized, split-localized and canonical orbitals, respectively, in the optimized VMC energy. Therefore it is seen that split-localized orbitals do give a significant improvement over canonical orbitals for this example, and so may be a reasonable compromise in cases where a fast CI convergence is needed. However their accuracy is still much lower compared to fully localized orbitals, and we observe this result consistently.

\subsection{Fixed-node approximation: examples}
\label{sec:fixed_node_examples}

\begin{table*}[t]
\begin{center}
\renewcommand{\arraystretch}{1.4}
{\footnotesize
\begin{tabular}{@{\extracolsep{4pt}}llcccc@{}}
\hline
\hline
Wave function & System & Active space & VMC error (mHa) & Fixed-node FCIQMC error (mHa) & Error removed ($\%$) \\
\hline
J-K$S_Z$GHF & H$_{14}$                & (14e,14o) & 1.71(2)   &  0.46(1)  & 73 \\
            & H$_{40}$                & (40e,40o) & 12.2(2)   &  2.4(2)   & 80 \\
            & TPA (C$_{16}$H$_{18}$)  & (16e,16o) & 0.426(8)  &  0.048(7) & 89 \\
            & TPA (C$_{28}$H$_{30}$)  & (28e,28o) & 1.36(2)   &  0.15(3)  & 89 \\
            & Coronene                & (24e,24o) & 6.72(3)   &  1.92(4)  & 71 \\
            & Hexacene                & (26e,26e) & 5.65(5)   &  1.58(3)  & 72 \\
            & Octacene                & (34e,34o) & 9.8(1)    &  2.8(1)   & 71 \\
            & BPEA                    & (30e,30o) & 5.50(6)   &  1.50(9)  & 73 \\
            & Ferrocene               & (10e,7o)  & 1.46(1)   &  0.95(5)  & 35 \\
            & Ferrocene               & (18e,15o) & 14.8(1)   &  9.53(3)  & 36 \\
            & Fe(P)                   & (32e,29o) & 9.24(6)   &  3.48(6)  & 62 \\
\hline
J-KAGP      & H$_{14}$                & (14e,14o) & 0.68(1)   &  0.183(9) & 73 \\
            & TPA (C$_{16}$H$_{18}$)  & (16e,16o) & 0.801(9)  &  0.17(1)  & 79 \\
            & Coronene                & (24e,24o) & 10.62(6)  &  3.52(5)  & 67 \\
            & BPEA                    & (30e,30o) & 4.71(4)   &  1.56(7)  & 67 \\
\hline
\hline
\end{tabular}
}
\end{center}
\caption{VMC and fixed-node FCIQMC errors in the ground-state energy for a variety of systems and active spaces, as defined in Section~\ref{sec:systems}. Errors are given relative to accurate DMRG benchmarks, except for Fe(P) where extrapolated SHCI is used.}
\label{tab:fixed_node}
\end{table*}

The above results demonstrate that the fixed-node approximation can be gradually switched off, provided a sufficient number of walkers are used, allowing stable sampling in the presence of a sign problem. However, it can be seen from Figure~\ref{fig:plateau_partial_node} that the walker population required is a significant fraction of that required for free propagation, which grows exponentially with system size. Therefore, for problems in large active spaces the walker population required becomes prohibitively large. Instead, we focus on the fixed-node approximation, using the Hamiltonian $\Hfn(\gamma=0.0)$ with the sign-flip potential applied.

We begin by applying this method to a variety of molecules, with results presented in Table~\ref{tab:fixed_node}. These systems include acene derivatives and trans-polyacetylene using full $\pi$-valence spaces, hydrogen chains in the minimal STO-6G basis, and ferrocene and Fe(P), with active spaces including $3d$ orbitals. These systems and active spaces were defined in Section~\ref{sec:systems}. We took $\epsilon = 10^{-6}$ Ha for coronene and BPEA, and $\epsilon = 10^{-8}$ Ha for all other systems.

First it can be noted that the J-K$S_Z$GHF wave function is typically very accurate as an active space solver. In most examples we have studied the final error in the ground state energy is less than $10$ mHa, and often much better. The percentage of error removed by fixed-node FCIQMC varies between systems and active spaces, but is often better than $70\%$, representing a significant improvement. There are a few outliers from this trend. In particular, the fixed-node approximation seems less accurate for ferrocene, removing around $35\%$ of error. It is not clear why this is, and fixed-node is seen to be effective for Fe(P), the other system studied containing $3d$ Fe orbitals. The J-K$S_Z$GHF wave function is also much less accurate in the larger ferrocene active space studied. We note that we encountered a few other instances where the fixed-node approximation was less effective, when using canonical or split localized orbitals, where the J-K$S_Z$GHF wave function is again less accurate.

We also performed some of these calculations using the J-KAGP wave function. We once again find this to usually be accurate as an active space solver. Compared to J-K$S_Z$GHF, variational energies from J-KAGP are more accurate for H$_{14}$ and BPEA, but slightly less accurate for C$_{16}$H$_{18}$ and coronene. Typically we find that J-K$S_Z$GHF wave functions are much easier to optimize than J-KAGP wave functions, requiring fewer iterations to reach the minimum energy. The percentage of error recovered by the fixed-node approximation in J-KAGP also seems to be slightly less in the examples studied, although it is not clear to what extent this is systematic. Overall, both wave functions are accurate and the fixed-node approach is commonly seen to remove $70\%$ of error or more, with only a few exceptions found.

\subsection{Fixed-node approximation: scaling}
\label{sec:fixed_node_scaling}

Next we look at two series of molecules in order to investigate scaling of both error and computational cost with system size. We look at trans-polyacetylene (TPA) molecules from C$_8$H$_{10}$ (8e,8o) to C$_{28}$H$_{30}$ (28e,28o), and acenes from napthalene (10e,10o) to octacene (34e,34o), using a full $\pi$-valence space for each example. An optimized J-K$S_Z$GHF wave function is used in each case.

\begin{figure}
\includegraphics{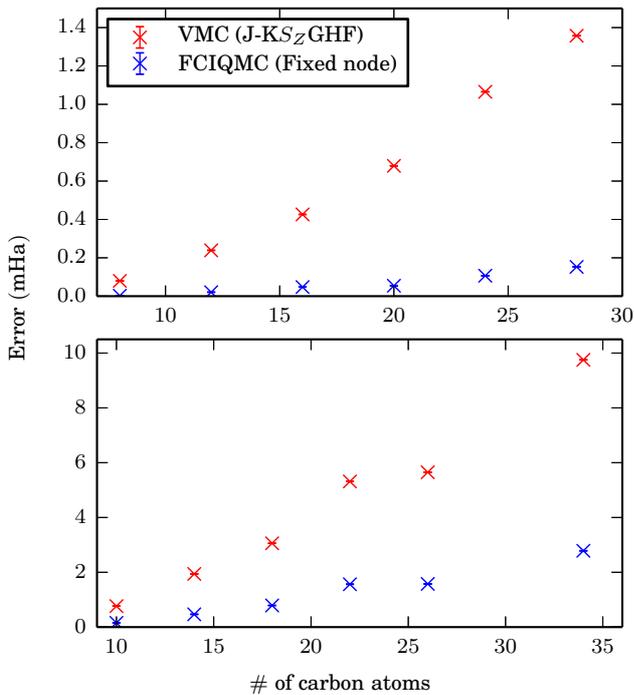}
\caption{Error in ground-state energies, comparing the optimized J-K$S_Z$GHF energy from VMC to the corresponding fixed-node FCIQMC energy. (a) trans-polyacetylene (TPA) from C$_8$H$_{10}$ (8e,8o) to C$_{28}$H$_{30}$ (28e,28o). (b) acenes from napthalene (10e,10o) to octacene (34e,34o). The active space in each case consists of all valence $\pi$ orbitals. The error is defined relative to DMRG benchmarks.}
\label{fig:error_scaling}
\end{figure}

Figure~\ref{fig:error_scaling} shows the error in the final estimates of the ground-state energy, taking $\gamma=0.0$. For TPA the J-K$S_Z$GHF wave function is extremely accurate, with an error of less than $1.4$ mHa for the largest active space (28e,28o). Nonetheless the fixed-node approximation improves this much further, removing about $89\%$ of this error. For acenes the error in J-K$S_Z$GHF grows roughly linearly with the system size to a largest value of $9.8$ mHa. The fixed-node approximation removes around $71\%$ of this error. Therefore the fixed-node approximation is found to be accurate and reliable up to challenging active space sizes.

\begin{figure}
\includegraphics{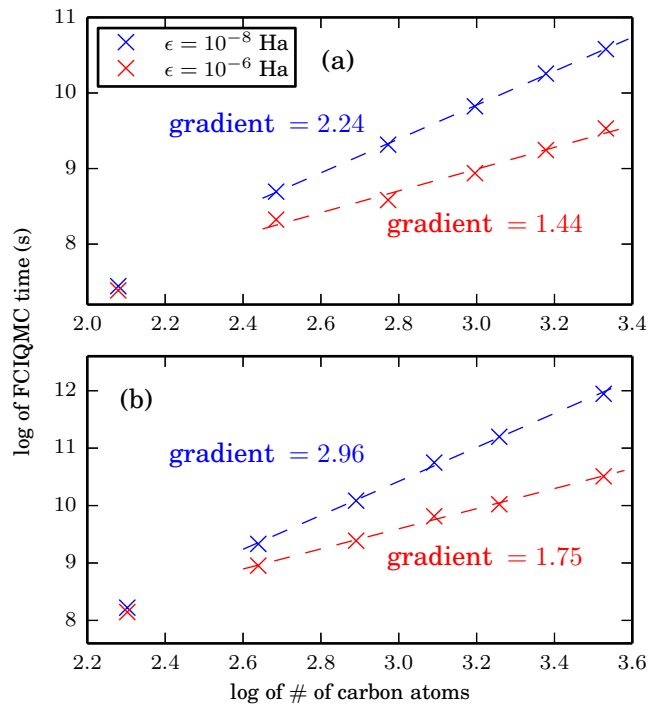}
\caption{Scaling of the wall time to perform $10^6$ iterations with fixed-node FCIQMC using $10^4$ walkers. (a) trans-polyacetylene (TPA) from C$_8$H$_{10}$ (8e,8o) to C$_{28}$H$_{30}$ (28e,28o). (b) acenes from napthalene (10e,10o) to octacene (34e,34o).}
\label{fig:time_scaling}
\end{figure}

We also investigate the scaling of computational cost with system and active space size in the fixed-node FCIQMC method. Figure~\ref{fig:time_scaling} looks at the same two systems as above. For both sets of systems we performed $10^6$ iterations with approximately $10^4$ walkers. For TPA a time step of $\Delta\tau = 4 \times 10^{-2}$ a.u. was used, while for the acenes $\Delta\tau = 2 \times 10^{-2}$ a.u. The number of iterations used is very large, leading to long simulation times and extremely small error bars. However this choice allows us to estimate error bars accurately for this scaling analysis. Fewer iterations would be performed in a typical calculation, resulting in much quicker simulations than those presented.

All simulations were performed on a single node consisting of two 8-core Xeon E5-2650 (2.6GHz) processors. As such the calculations did not use significant parallel computing resources, although the method is scalable to large numbers of CPU cores, as for the traditional FCIQMC method\cite{Booth2014}.

We investigated two values of the parameter $\epsilon$, which determines the Hamiltonian element cutoff when calculating the local energy and spin-flip potential, as in Eq.~(\ref{eq:local_E_HB}). Specifically we took $\epsilon=10^{-8}$ Ha and $\epsilon=10^{-6}$ Ha.

Figure~\ref{fig:time_scaling} shows a log-log plot of the total wall time against the number of carbon atoms, $n_c$ (which is also the number of active space orbitals). We aimed to use $10^4$ walkers in each calculation, but there is small fluctuation in the average value, by up to $10\%$, that is hard to control. We therefore plot $T \times 10^{4} / N_{\mathrm{av}}$ to correct for this issue, where $N_{\mathrm{av}}$ is the average walker population over the entire simulation. It is seen that the approach has a low polynomial scaling. For TPA and acenes the asymptotic scaling appears to be around $T \sim \mathcal{O}(n_c^{2.24})$ and $T \sim \mathcal{O}(n_c^{2.96})$ respectively, with $\epsilon = 10^{-8}$ Ha. When $\epsilon = 10^{-6}$ Ha this scaling appears to reduce to $T \sim \mathcal{O}(n_c^{1.44})$ and $T \sim \mathcal{O}(n_c^{1.75})$, demonstrating very low scaling in the total computational time to perform fixed-node FCIQMC. Given the high accuracy observed with fixed-node J-K$S_Z$GHF for these challenging CASCI problems, this is very encouraging.

We do not find any significant reduction in accuracy from choosing $\epsilon=10^{-6}$ Ha compared to $\epsilon=10^{-8}$ Ha. For example, for C$_{28}$H$_{30}$ we found the discrepancy to be $0.091(44)$ mHa, which is essentially negligible compared to the reduction in simulation time. We also ran the TPA calculations with $\epsilon=10^{-4}$ Ha. Here the scaling of the wall time is reduced further, but the accuracy is significantly worsened. The discrepancy with the $\epsilon=10^{-8}$ Ha result for C$_{28}$H$_{30}$ is $4.5$ mHa. Therefore we find $\epsilon=10^{-6}$ Ha to be a sensible trade-off for practical calculations.

It is also important to look at the scaling of the final statistical error on the energy estimate. Given the correlated nature of FCIQMC data, we estimate this error by performing a reblocking analysis\cite{Flyvbjerg1989}. To ensure that the error chosen from the scaling analysis is objective, we use an automated criteria to choose the optimal block length. The criteria used is that suggested by Lee \emph{et al}.,\cite{Lee2011} implemented in the pyblock package, which we use for this analysis\cite{pyblock}. Specifically, the block size $B$ is chosen as the smallest block for which for which $B^3 > 2n\eta_{\textrm{err}}^4(B)$ is satisfied. Here, $n$ is the total number of samples, and $\eta_{\textrm{err}}(B)$ is an estimate of the square root of the correlation length from block length $B$. This is obtained as $\eta_{\textrm{err}}(B) = \sigma_{\textrm{corr.}}(B)/\sigma_{\textrm{uncorr.}}$, where $\sigma_{\textrm{corr.}}(B)$ is the error estimate from block length $B$ and $\sigma_{\textrm{uncorr.}}$ is the error obtained from an uncorrelated analysis (with $B=1$). We refer to ref~\onlinecite{Lee2011} for details. This criteria for the block length is based on that of Wolff\cite{Wolff2004}, which aims to balance the systematic and statistical error in the error estimate. As discussed above, a large number of iterations and a large time step were taken to reduce this uncertainty.

Estimates of statistical error for calculations from Figure~\ref{fig:time_scaling} (with $\epsilon=10^{-6}$ Ha) are shown in Table~\ref{tab:error_scaling}. To correct for small differences in $N_{\mathrm{av}}$, the values presented are $\sigma \times \sqrt{N_{\mathrm{av}}/10^4}$. As can be seen, using $10^6$ iterations and with accurate J-K$S_Z$GHF wave functions, the error is extremely small and typically at least an order of magnitude smaller than the fixed-node error. A power law scaling is not quite so clear in this case, but discarding the first data point gives an approximate scaling for the error of $\sigma \sim \mathcal{O}(n_c^{2.6})$ and $\sigma \sim \mathcal{O}(n_c^{2.5})$ for TPA and acenes, respectively. Because the error decreases with the simulation time as $T^{-1/2}$, one measure for the total cost is $\eta = T \times \sigma^2$. This gives a higher total scaling of $\eta \sim \mathcal{O}(n_c^{6.7})$ for both systems. This metric accounts for having to perform more iterations to achieve a fixed statistical error with increasing system size. However, in practice the error bar is so small that these extra iterations are not required, and so this worse scaling is not observed for this range of systems, although will be eventually.

The correlation length is another important factor in the scaling of QMC methods, which often determines the number of iterations which must be performed. From a practical point of view, this can be determined by the requirements for an accurate reblocking analysis. In both TPA and acenes, this block length is found to be independent of system size for the range of systems studied. The automated procedure chooses a block length of $2^{11}$ iterations for each TPA molecule, with a fixed simulation time step of $\Delta\tau=0.04$ a.u., and a block length of $2^{12}$ for each linear acene, with $\Delta\tau=0.02$ a.u. Only blocks of length $2^n$ were considered. Therefore from a practical point of view we find no contribution to the scaling from an increasing correlation length. We can also look at the $\eta_{\mathrm{err}}^2$ as an estimate of correlation length. This is found to vary between $52$ and $62$ for the TPA examples, and $88$ to $133$ for acene examples, and only a slight upward trend is clear. However, the scaling of the correlation length is likely system dependent, and higher scaling may be observed in other examples.

Numerical data plotted in Figure~\ref{fig:time_scaling} is given in the Supporting Information, together with additional data at $\epsilon=10^{-4}$ Ha and estimates of $\eta_{\mathrm{err}}^2$.

Therefore, for the range of systems studied here one can take a constant number of iterations in fixed-node FCIQMC, such that the scaling is the increase in total wall time observed in Figure~\ref{fig:time_scaling}. For very large active spaces the increase in statistical error will likely become the limiting factor, and the scaling to achieve a constant error will become higher. Other effects may dominate for very large systems\cite{Nemec2010}.

\begin{table}[t]
\begin{center}
{\footnotesize
\begin{tabular}{@{\extracolsep{4pt}}llc@{}}
\hline
\hline
System type & $\#$ of carbon atoms & Statistical error (mHa) \\
\hline
TPA    & 8    &  $9.7 \times 10^{-4}$ \\ 
       & 12   &  $3.4 \times 10^{-3}$ \\
       & 16   &  $6.9 \times 10^{-3}$ \\
       & 20   &  $1.2 \times 10^{-2}$ \\
       & 24   &  $2.1 \times 10^{-2}$ \\
       & 28   &  $3.1 \times 10^{-2}$ \\
\hline
Acenes & 10   &  $4.2 \times 10^{-3}$\\ 
       & 14   &  $1.2 \times 10^{-2}$\\
       & 18   &  $2.1 \times 10^{-2}$\\
       & 22   &  $4.0 \times 10^{-2}$\\
       & 26   &  $5.5 \times 10^{-2}$\\
       & 34   &  $1.1 \times 10^{-1}$\\
\hline
\hline
\end{tabular}
}
\end{center}
\caption{Scaling of the statistical error on the energy estimate from fixed-node FCIQMC, as a function of the number of carbon atoms, $n_c$. The systems are trans-polyacetylene molecules from C$_8$H$_{10}$ (8e,8o) to C$_{28}$H$_{30}$ (28e,28o) and acenes from napthalene (10e,10o) to octacene (34e,34o).}
\label{tab:error_scaling}
\end{table}

All fixed-node FCIQMC calculations must be preceded by optimization of the VMC wave function. We find this to be the more expensive step for larger systems, due to the increased difficulty in fully optimizing the wave function. Performing a constant number of iterations with VMC has a low polynomial scaling, which we observe to be similar to that for fixed-node FCIQMC in Figure~\ref{fig:time_scaling}. However, the number of optimization steps to reach a desired accuracy threshold also increases. In particular, a long tail is observed for larger active spaces, perhaps suggesting a vanishing gradient problem, as observed when training artificial neural networks\cite{Hochreiter1998}. To give an example for BPEA (30e,30o), we consider optimizing the J-K$S_Z$GHF wave function with the following parameters: $1000$ samples per process per optimization step, with $\epsilon = 10^{-6}$ Ha and an AMSGrad step size of $0.01$. After $1000$, $6000$ and $12000$ AMSGrad optimization steps, the J-K$S_Z$GHF energy had a convergence error of approximately $4.9$ mHa, $1.1$ mHa and $0.3$ mHa, taking $0.94$, $5.66$ and $11.32$ hours to perform, respectively. The same $16$-core node as above was used. The subsequent fixed-node FCIQMC calculation took $2.1$ hours to perform $3 \times 10^{5}$ iterations with $10^{4}$ walkers on the same node, again setting $\epsilon = 10^{-6}$ Ha, which gave a final statistical error of $0.1$ mHa. In this case the fixed-node FCIQMC simulation is seen to be quicker to perform, which we find to be typical for larger active spaces, depending on the degree of convergence achieved in the VMC optimization. It may be sensible to use an approach such as the linear method\cite{Umrigar2007} to perform the remainder of the optimization, once the vicinity of the global minimum has been approximately reached, as has been suggested recently\cite{Otis2019, Sabzevari2020}.

\section{Discussion and comparison to related methods}
\label{sec:comparison}

The fixed and partial-node methods presented build upon the FCIQMC approach, sharing a common walker propagation method to sample the wave function. Here we briefly discuss and compare these methods, as related approaches for obtaining high-accuracy approximations in active spaces beyond the reach of exact FCI. This area has seen important developments in recent years. This is emphasised in recent studies, including a comparison of several state-of-the-art methods applied to benzene\cite{Eriksen2020} and subsequent investigations\cite{Lee2020, Loos2020}. These methods include DMRG\cite{White1992, Chan2002, Olivares2015}, selected CI methods\cite{Tubman2016, Holmes2016_2, Liu2016}, full coupled cluster reduction (FCCR)\cite{Enhua2018}, many-body expanded FCI (MBE-FCI)\cite{Eriksen2017} and FCIQMC based on its adaptive-shift variant\cite{Ghanem2019} and cluster-analysis-driven methods\cite{Deustua2018}.

Comparing the fixed-node method to existing FCIQMC approaches, there is a significant difference in the nature of the approximation applied. FCIQMC studies have primarily applied the initiator approach to ameliorate the fermion sign problem\cite{Cleland2010}. This approximation has also been used within related approaches, such as model space quantum Monte Carlo (MSQMC)\cite{Ten-no2013}. In the standard initiator approximation, spawning events are cancelled if they occur to an unoccupied determinant, unless the parent population exceeds a minimum threshold. This allows all spawning from the more important determinants, but restricts spawning from the determinants with few walkers, effectively setting the corresponding Hamiltonian elements to $0$. This limits the Hamiltonian beyond the space of highly-occupied determinants. There is no trial wave function in this approach, but its accuracy depends on the walker population used. The FCI solution can be converged upon with increasing walker population. In contrast, the accuracy of the fixed-node approximation is independent of the walker population, and only depends on the accuracy of the trial wave function used. A drawback of this approach is the need to perform a prior VMC optimization of the trial wave function, which is non-trivial in general. However, there are a wide range of trial wave functions available, many of which are appropriate for strongly-correlated systems. This approach is therefore very general, and may be accurate in situations where the initiator method requires large walker populations.

As an example, we compare our fixed-node results from Section~\ref{sec:fixed_node_examples} to those from a recent initiator FCIQMC study\cite{Blunt2019_2}. We consider results for the BPEA molecule, as the same active space is studied as in ref~\citenum{Blunt2019_2}. This study used the i-FCIQMC(SCI) method, where the initiator space is augmented with determinants obtained from a prior selected CI calculation. As described above, a comparison between the two approaches is challenging because the accuracy of the i-FCIQMC approach depends on the walker population, whereas the fixed-node approximation does not. However, it can be seen from ref~\citenum{Blunt2019_2} that a walker population of over $10^6$ is required in order to achieve better than $2$ mHa in the variational ground-state energy estimate. In the fixed-node approach presented here, the final error is $1.50(9)$ mHa and $1.56(7)$ mHa for the J-K$S_Z$GHF and J-KAGP wave functions respectively, using $10^4$ walkers for both. Therefore the memory and time requirements are much lower in the fixed-node approach in this example. Another benefit is the low-polynomial scaling observed in Section~\ref{sec:fixed_node_scaling}, compared to that in the initiator method\cite{Cleland2011}. A drawback is that it is much more challenging to converge the fixed-node approach to the FCI limit (although trial wave functions using CI expansions have been considered within determinant-space VMC\cite{Mahajan2020}).

Moreover, recent improvements to the initiator approach have been made. These include the adaptive-shift method\cite{Ghanem2019}, and corrections from perturbation theory\cite{Blunt2018} and the coupled electron pair approximation within initiator MSQMC\cite{Ten-no2017}, each of which can often remove a significant fraction of initiator error. In particular, these approaches account for a significant fraction of size consistency error. It is not feasible to include a detailed comparison with all of these methods, but the approximations involved remain very distinct from the fixed-node method and the wave functions optimized within VMC. Indeed, because the fixed-node approach makes use of a trial wave function with a non-linear parameterization, the approximation applied perhaps shares more similarities with the DMRG method, which takes a matrix-product state representation of the wave function. It also shares some of its drawbacks, including difficulty accurately treating dynamical correlation in large virtual spaces, and some of the approaches to address this within DMRG may be useful\cite{Guo2018_1}. Ultimately, we hope that the ability of this method to work with a wide variety of trial wave functions, with low-polynomial scaling and without dependence on the walker population, makes it a powerful approach with many directions to explore.

\section{Conclusion}

We have presented a study of fixed and partial-node approximations for complete active space problems in \emph{ab initio} systems, using symmetry-projected Jastrow mean-field wave functions and the FCIQMC method to perform sampling. This leads to an FCIQMC method with a low polynomial scaling computational cost in active space size, without a significant increase in associated error.

The annihilation of walkers in the FCIQMC method allows the fixed-node approximation to be partially lifted while maintaining stable sampling. For small enough systems this allows us to converge to the exact limit. However, the walker population required to achieve this grows quickly, such that this approach is not viable for very large active spaces. In a basis of localized orbitals, importance sampling is found to slightly reduce the sign problem severity. When using canonical or split-localized orbitals, importance sampling generally has a negative effect on both the sign problem severity and statistics.

In contrast, the fixed-node approximation in FCIQMC is found to be accurate and scalable. With the symmetry-projected Jastrow mean-field wave functions, we often achieve an accuracy within a few millihartrees from the exact ground-state energy. Use of such trial wave functions within FCIQMC has been avoided in the past due to the potentially prohibitive cost of evaluating the local energy, requiring a summation over all connections for each walker. However, the heat bath approach is found to largely remove such issues. By setting the heat bath threshold appropriately, we find that the scaling of total wall time with system size is very low. More generally, these trial wave functions are useful in FCIQMC beyond the fixed-node approximation, particularly in the mixed energy estimator.

Our results also further demonstrate the accuracy of symmetry-projected Jastrow-GHF and AGP wave functions for complete active space problems. Given that fixed-node FCIQMC improves the accuracy of these wave functions and is often faster to perform than the preceding VMC optimization, we believe that this is an effective and practical approach.

Including dynamical correlation beyond these CAS-based calculations is the next important step. We have recently demonstrated a VMC-based approach for multi-reference perturbation theory using selected CI wave functions\cite{Mahajan2019_2, Blunt2020}, which if extended to VMC and fixed-node FCIQMC could provide a polynomial scaling approach for treating both dynamical and static correlation. Additionally, we hope that this work will provide the starting point for improved approximations in FCIQMC beyond the traditional fixed-node approach, and other improvements to the FCIQMC method more generally.

\section{Supporting Information}
The Supporting Information includes numerical data for values plotted in Figures~\ref{fig:plateau_partial_node} to \ref{fig:time_scaling}, additional results for scaling of computational cost with system size, and example J-K$S_Z$GHF calculations in larger basis sets. Geometries are presented for all systems studied.

\begin{acknowledgments}
We thank Sandeep Sharma for discussions and comments on this manuscript. We are grateful to St John's College, Cambridge for funding this work through a Research Fellowship. This study made use of the CSD3 Peta4-Skylake CPU cluster.
\end{acknowledgments}


\begin{thebibliography}{95}
\bibitem [{fn1()}]{fn1}%
  \BibitemOpen
  \href@noop {} {}\bibinfo {note} {We have performed preliminary calculations
  to demonstrate this point, which are included in the Supporing
  Information.}\BibitemShut {Stop}%
\bibitem [{fn2()}]{fn2}%
  \BibitemOpen
  \href@noop {} {}\bibinfo {note} {This is an important feature of CI
  expansions. They are not invariant to orbital rotations, and will converge
  much more slowly in certain orbital basis sets. This can be seen clearly by
  performing selected CI calculations. For example, we perform HCI on H$_{14}$
  (STO-6G, $R=2$ $a_0$) with a threshold of $\epsilon=10^{-4}$ Ha. In a basis
  of canonical orbitals, the expansion consists of $1.5 \times 10^5$
  determinants with a final error of $2.3$ mHa in the ground-state energy. In
  localized orbitals, the same calculations leads to a HCI expansion of $1.3
  \times 10^6$ determinants and an error of $21$ mHa in the ground-state
  energy.}\BibitemShut {Stop}%
\end{thebibliography}
%

\end{document}


\maketitle

\section{Additional data}

Data is presented for results in the main article. Table 1 presents the numerical plateau heights that are plotted in Figure 2. In Table 2, data is given for partial node simulations presented in Figure 3. These include the numerical values plotted, and also relevant FCIQMC simulation parameters. The benchmark energies obtained from the DMRG method are also given. Table 3 presents the values plotted in Figure 4 of the main article. Tables 4 and 5 present data from the scaling analysis, including walker populations and correlation length estimates. For trans-polyacetylene, we also include data at $\epsilon=10^{-4}$ Ha, which is not included in the main article.

Tables 6 and 7 present data to demonstrate that the J-K$S_Z$GHF wave function loses accuracy in large orbital basis sets.

\subsection{Plateau heights for H$_{14}$}

\begin{table*}
\begin{center}
\caption{Plateau heights plotted in Figure 2, using partial-node FCIQMC, as a function of the fixed-node factor $\gamma$. The system in H$_{14}$ in a STO-6G basis at $R=2$ $a_0$.}

{\footnotesize
\begin{tabular}{@{\extracolsep{4pt}}lc@{}}
\hline
\hline
$\gamma$ & Walker population plateau height \\
\hline
 0.0  &  0      \\
-0.2  &  7800   \\
-0.4  &  33933  \\
-0.6  &  97100  \\
-0.8  &  235538 \\
-1.0  &  561560 \\
\hline
\hline
\end{tabular}
}

\end{center}
\end{table*}

\newpage

\subsection{Partial-node results}

\begin{table*}
\begin{subtable}{\textwidth}\centering
\caption{H$_{14}$ in a STO-6G basis and $R=2$ $a_0$. The DMRG benchmark ($M=1000$) is -7.5422705 Ha. The time step is 0.001 a.u. in each case.}
{\footnotesize
\begin{tabular}{@{\extracolsep{4pt}}lcccc@{}}
\hline
\hline
& \multicolumn{2}{c}{With $\VSF$} & \multicolumn{2}{c}{Without $\VSF$} \\
\cline{2-3} \cline{4-5}
$\gamma$ & Energy (Ha) & Walker pop. & Energy (Ha) & Walker pop. \\
\hline
 0.0  &  -7.541747(8) & $2.3 \times 10^5$ &  -7.54148(1)  & $2.3 \times 10^5$ \\ 
-0.2  &  -7.54182(1)  & $2.3 \times 10^5$ &  -7.54157(1)  & $2.3 \times 10^5$ \\ 
-0.4  &  -7.54190(2)  & $2.3 \times 10^5$ &  -7.54171(2)  & $2.3 \times 10^5$ \\ 
-0.6  &  -7.54200(2)  & $2.3 \times 10^5$ &  -7.54188(2)  & $4.6 \times 10^5$ \\ 
-0.8  &  -7.54215(1)  & $9.2 \times 10^5$ &  -7.54207(3)  & $9.2 \times 10^5$ \\ 
-1.0  &  -7.54225(5)  & $1.2 \times 10^6$ &  -7.54226(3)  & $1.2 \times 10^6$ \\ 
\hline
\hline
\end{tabular}
}
\end{subtable}

\vspace{6mm}

\begin{subtable}{\textwidth}\centering
\caption{Anthracene (14e,14o). The DMRG benchmark ($M=1000$) is -535.9946890 Ha. The time step is 0.001 a.u. in each case.}
{\footnotesize
\begin{tabular}{@{\extracolsep{4pt}}lcccc@{}}
\hline
\hline
& \multicolumn{2}{c}{With $\VSF$} & \multicolumn{2}{c}{Without $\VSF$} \\
\cline{2-3} \cline{4-5}
$\gamma$ & Energy (Ha) & Walker pop. & Energy (Ha) & Walker pop. \\
\hline
 0.0  & -535.99423(2)  & $2.1 \times 10^4$ &  -535.993810(7)  & $2.1 \times 10^5$ \\ 
-0.2  & -535.994250(5) & $2.1 \times 10^5$ &  -535.99399(1)   & $2.1 \times 10^5$ \\ 
-0.4  & -535.994316(7) & $2.1 \times 10^5$ &  -535.99416(2)   & $2.1 \times 10^5$ \\ 
-0.6  & -535.99441(1)  & $2.1 \times 10^5$ &  -535.99432(3)   & $2.1 \times 10^5$ \\ 
-0.8  & -535.994490(6) & $6.2 \times 10^5$ &  -535.99449(2)   & $6.2 \times 10^5$ \\ 
-0.9  & -535.99456(1)  & $6.2 \times 10^5$ &  -               & - \\
-1.0  & -535.99469(6)  & $8.3 \times 10^5$ &  -535.99465(4)   & $8.3 \times 10^5$ \\ 
\hline
\hline
\end{tabular}
}
\end{subtable}

\vspace{6mm}

\begin{subtable}{\textwidth}\centering
\caption{Trans-polyacetylene C$_{12}$H$_{14}$ (12e,12o). The DMRG benchmark ($M=1000$) is -462.470003 Ha.}
{\footnotesize
\begin{tabular}{@{\extracolsep{4pt}}lcccccc@{}}
\hline
\hline
& \multicolumn{3}{c}{With $\VSF$} & \multicolumn{3}{c}{Without $\VSF$} \\
\cline{2-4} \cline{5-7}
$\gamma$ & Energy (Ha) & Walker pop. & Time step & Energy (Ha) & Walker pop. & Time step \\
\hline
 0.0  & -462.467868(2)   & $2.0 \times 10^4$ & 0.0005 & -462.46739(1)   & $8.0 \times 10^4$ & 0.0005 \\
-0.1  & -462.467940(1)   & $2.0 \times 10^4$ & 0.0005 & -462.468843(9)  & $8.0 \times 10^4$ & 0.0005 \\
-0.2  & -462.468023(2)   & $2.0 \times 10^4$ & 0.0005 & -462.469441(7)  & $8.0 \times 10^4$ & 0.0005 \\
-0.4  & -462.468225(3)   & $2.0 \times 10^4$ & 0.0005 & -462.469788(8)  & $1.0 \times 10^5$ & 0.0005 \\
-0.6  & -462.468511(4)   & $2.0 \times 10^4$ & 0.0005 & -462.46987(1)   & $1.5 \times 10^5$ & 0.0005 \\
-0.8  & -462.468928(1)   & $2.0 \times 10^5$ & 0.0005 & -462.469915(3)  & $2.5 \times 10^5$ & 0.0005 \\
-0.9  & -462.46927531(6) & $4.0 \times 10^5$ & 0.0005 & -462.4699570(9) & $4.0 \times 10^5$ & 0.001  \\
-1.0  & -462.470003(1)   & $6.0 \times 10^5$ & 0.001  & -462.470005(1)  & $6.0 \times 10^5$ & 0.001  \\
\hline
\hline
\end{tabular}
}
\end{subtable}
\caption{Results from partial-node FCIQMC, Figure $3$ in the main article. Average walker populations are given to 2 s.f. Walker populations and time steps were not optimized.}
\end{table*}

\newpage


\subsection{Fixed-node results: energy scaling}

\begin{table*}
\begin{subtable}{\textwidth}\centering
\caption{Trans-polyacetylene molecules from C$_8$H$_{10}$ (8e,8o) to C$_{28}$H$_{30}$ (28e,28o)}
{\footnotesize
\begin{tabular}{@{\extracolsep{4pt}}lccc@{}}
\hline
\hline
 & \multicolumn{3}{c}{Ground-state energy (Ha)} \\
\cline{2-4}
$\#$ of carbon atoms, $n_c$ & VMC (J-K$S_Z$GHF)& Fixed-node FCIQMC & DMRG benchmark \\
\hline
8  & -308.695267(3)  & -308.695344(1)  & -308.6953466  \\
12 & -462.469764(5)  & -462.469982(3)  & -462.4700031  \\
16 & -616.243875(8)  & -616.244254(7)  & -616.2443013  \\
20 & -770.01746(1)   & -770.01808(1)   & -770.0181351  \\
24 & -923.79016(2)   & -923.79112(2)   & -923.7912282  \\
28 & -1077.56284(2)  & -1077.56405(3)  & -1077.5642004 \\
\hline
\hline
\end{tabular}
}
\end{subtable}

\vspace{1cm}

\begin{subtable}{\textwidth}\centering
\caption{Acenes from naphthalene (10e,10o) to octacene (34e,34o)}
{\footnotesize
\begin{tabular}{@{\extracolsep{4pt}}lccc@{}}
\hline
\hline
 & \multicolumn{3}{c}{Ground-state energy (Ha)} \\
\cline{2-4}
$\#$ of carbon atoms, $n_c$ & VMC (J-K$S_Z$GHF)& Fixed-node FCIQMC & DMRG benchmark \\
\hline
10  & -383.349363(8) & -383.34997(1) & -383.3501284  \\
14  & -535.99275(1)  & -535.99422(2) & -535.9946890  \\
18  & -688.63348(2)  & -688.63575(4) & -688.6365360  \\
22  & -841.27183(3)  & -841.27558(4) & -841.2771446  \\
26  & -993.91138(5)  & -993.91546(3) & -993.9170322  \\
34  & -1299.1860(1)  & -1299.1930(1) & -1299.1957740 \\
\hline
\hline
\end{tabular}
}
\end{subtable}
\caption{Estimates of the ground-state energy from VMC (using an optimized J-K$S_Z$GHF wave function), fixed-node FCIQMC, and DMRG benchmarks. This data is plotted in Figure $4$ in the main article.}
\end{table*}

\newpage


\subsection{Fixed-node results: cost scaling for trans-polyacetylene}

\begin{table*}
\begin{subtable}{\textwidth}\centering
\caption{Results with heat bath threshold $\epsilon = 10^{-4}$ Ha.}
{\footnotesize
\begin{tabular}{@{\extracolsep{4pt}}lcccc@{}}
\hline
\hline
$\#$ of carbon atoms, $n_c$ & Wall time (s) & Statistical error (mHa) & $\eta_{\mathrm{err}}^2$ & Average walker pop. \\
\hline
8   &  1627  & $1.0 \times 10^{-3}$  & 53.7 & 10188 \\
12  &  2951  & $3.7 \times 10^{-3}$  & 59.2 & 9950  \\
16  &  3278  & $6.9 \times 10^{-3}$  & 60.6 & 9963  \\
20  &  4582  & $1.2 \times 10^{-2}$  & 48.3 & 10021 \\
24  &  5920  & $2.1 \times 10^{-2}$  & 57.8 & 9942  \\
28  &  7664  & $3.2 \times 10^{-2}$  & 59.7 & 9816  \\
\hline
\hline
\end{tabular}
}
\end{subtable}

\vspace{1cm}

\begin{subtable}{\textwidth}\centering
\caption{Results with heat bath threshold $\epsilon = 10^{-6}$ Ha.}
{\footnotesize
\begin{tabular}{@{\extracolsep{4pt}}lcccc@{}}
\hline
\hline
$\#$ of carbon atoms, $n_c$ & Wall time (s) & Statistical error (mHa) & $\eta_{\mathrm{err}}^2$ & Average walker pop. \\
\hline
8   &  1653   & $9.6 \times 10^{-4}$  & 52.6  &  10234  \\
12  &  4040   & $3.4 \times 10^{-3}$  & 55.0  &  9801   \\
16  &  5350   & $6.9 \times 10^{-3}$  & 54.0  &  10013  \\
20  &  7629   & $1.2 \times 10^{-2}$  & 53.0  &  10001  \\
24  &  10128  & $2.1 \times 10^{-2}$  & 57.9  &  9785   \\
28  &  13446  & $3.1 \times 10^{-2}$  & 61.8  &  9760   \\
\hline
\hline
\end{tabular}
}
\end{subtable}

\vspace{1cm}

\begin{subtable}{\textwidth}\centering
\caption{Results with heat bath threshold $\epsilon = 10^{-8}$ Ha.}
{\footnotesize
\begin{tabular}{@{\extracolsep{4pt}}lcccc@{}}
\hline
\hline
$\#$ of carbon atoms, $n_c$ & Wall time (s) & Statistical error (mHa) & $\eta_{\mathrm{err}}^2$ & Average walker pop. \\
\hline
8   & 1745   &  $9.9 \times 10^{-4}$ & 57.6 & 10207 \\
12  & 5954   &  $3.4 \times 10^{-3}$ & 63.1 & 9977  \\
16  & 11048  &  $7.4 \times 10^{-3}$ & 52.3 & 9940  \\
20  & 18265  &  $1.2 \times 10^{-2}$ & 52.1 & 9894  \\
24  & 27847  &  $2.1 \times 10^{-2}$ & 55.5 & 9785  \\
28  & 40410  &  $3.0 \times 10^{-2}$ & 61.8 & 10264 \\
\hline
\hline
\end{tabular}
}
\end{subtable}
\caption{Results for the scaling of FCIQMC computational cost with system size, for trans-polyacetylene molecules from C$_8$H$_{10}$ (8e,8o) to C$_{28}$H$_{30}$ (28e,28o). $10^6$ iterations were performed for each calculation.}
\end{table*}

\newpage


\subsection{Fixed-node results: cost scaling for acenes}

\begin{table*}
\begin{subtable}{\textwidth}\centering
\caption{Results with heat bath threshold $\epsilon = 10^{-6}$ Ha.}
{\footnotesize
\begin{tabular}{@{\extracolsep{4pt}}lcccc@{}}
\hline
\hline
$\#$ of carbon atoms, $n_c$ & Wall time (s) & Statistical error (mHa) & $\eta_{\mathrm{err}}^2$ & Average walker pop. \\
\hline
10  & 3441   &  $4.2 \times 10^{-3}$  & 88.1  &  9973  \\
14  & 7756   &  $1.2 \times 10^{-2}$  & 117.9 &  10028 \\
18  & 13257  &  $2.0 \times 10^{-2}$  & 106.9 &  11106 \\
22  & 18058  &  $4.0 \times 10^{-2}$  & 120.6 &  9872  \\
26  & 22692  &  $5.5 \times 10^{-2}$  & 133.3 &  10066 \\
34  & 36529  &  $1.1 \times 10^{-1}$  & 131.4 &  9967  \\
\hline
\hline
\end{tabular}
}
\end{subtable}

\vspace{1cm}

\begin{subtable}{\textwidth}\centering
\caption{Results with heat bath threshold $\epsilon = 10^{-8}$ Ha.}
{\footnotesize
\begin{tabular}{@{\extracolsep{4pt}}lcccc@{}}
\hline
\hline
$\#$ of carbon atoms, $n_c$ & Wall time (s) & Statistical error (mHa) & $\eta_{\mathrm{err}}^2$ & Average walker pop. \\
\hline
10  & 3739   &  $4.4 \times 10^{-3}$  & 94.3  & 10046 \\
14  & 11293  &  $1.2 \times 10^{-2}$  & 122.9 & 10003 \\
18  & 26402  &  $2.0 \times 10^{-2}$  & 112.8 & 11009 \\
22  & 46347  &  $4.0 \times 10^{-2}$  & 124.3 & 9959  \\
26  & 73223  &  $5.2 \times 10^{-2}$  & 122.3 & 10036 \\
34  & 155267 &  $1.1 \times 10^{-1}$  & 144.4 & 10075 \\
\hline
\hline
\end{tabular}
}
\end{subtable}
\caption{Results for the scaling of FCIQMC computational cost with system size for acenes from napthalene (10e,10o) to octacene (34e,34o). $10^6$ iterations were performed for each calculation.}
\end{table*}

\newpage


\subsection{J-K$S_Z$GHF error for increasing basis set size}

It is claimed in the main article that the wave functions used have low accuracy for treating dynamical correlation in large orbital basis sets. Here we include data to demonstrate this. We use the J-K$S_Z$GHF wave function to study H$_{10}$ and N$_2$ in orbital basis sets of increasing size. The VMC method is used to optimize this wave function and estimate the ground-state energy, as described in the main article. It is seen that the ground-state energy estimate worsens considerably for larger basis sets. However, the J-K$S_Z$GHF wave function can be very accurate as an active space solver in CASSCI or CASSCF methods, as demonstrated by results in the main article. Dynamical correlation can then be included by other standard approaches, such as multireference perturbation theory. We have suggested one such approach to perform this using VMC recently\cite{Blunt2020}.

The benchmarks here were performed using the semi-stochastic heat-bath CI (SHCI) method\cite{Holmes2016_2, Sharma2017}. For STO-6G and 6-31G basis sets, this provided a near-exact FCI energy without extrapolation. For cc-pVDZ and cc-pVTZ basis sets, a simple quadratic extrapolation was performed to obtain an accurate FCI energy estimate, as described in previous SHCI studies\cite{Sharma2017, Smith2017}.

\begin{table*}
\begin{center}
\caption{Errors in ground-state energies from the J-K$S_Z$GHF wave function with different orbital basis sets, for H$_{10}$ with R=$2.0$ $a_0$.}

{\footnotesize
\begin{tabular}{@{\extracolsep{4pt}}lcc@{}}
\hline
\hline
Basis set & Active space & Energy error (mHa) \\
\hline
STO-6G    & (10e,10o)    &  0.9  \\
6-31G     & (10e,20o)    &  2.8  \\
cc-pVDZ   & (10e,50o)    &  58.1(2) \\
\hline
\hline
\end{tabular}
}

\end{center}
\end{table*}

\begin{table*}
\begin{center}
\caption{Errors in ground-state energies from the J-K$S_Z$GHF wave function with different orbital basis sets, for N$_2$ with R=$2.0$ $a_0$.}

{\footnotesize
\begin{tabular}{@{\extracolsep{4pt}}lcc@{}}
\hline
\hline
Basis set & Active space & Energy error (mHa) \\
\hline
STO-6G    & (14e,10o)    & 0.9      \\
6-31G     & (14e,18o)    & 12.3     \\
cc-pVDZ   & (14e,28o)    & 51.2     \\
cc-pVTZ   & (14e,60o)    & 123(1)   \\
\hline
\hline
\end{tabular}
}

\end{center}
\end{table*}

\newpage


\begin{table}
\begin{flushleft}
{\large \textbf{Geometries (\AA)}}

\vspace{5mm}

{\large \textbf{Acenes}}

\vspace{5mm}

\texttt{Napthalene}

\vspace{5mm}

\footnotesize{
\texttt{
\begin{tabular}{@{}lrrr@{}}
H &   2.337637658560 &  4.693101235084 &  0.000000000000 \\
H &  -2.337637658560 & -4.693101235084 &  0.000000000000 \\
H &  -2.337637658560 &  4.693101235084 &  0.000000000000 \\
H &   2.337637658560 & -4.693101235084 &  0.000000000000 \\
H &   6.364924545043 &  2.349824188415 &  0.000000000000 \\
H &  -6.364924545043 & -2.349824188415 &  0.000000000000 \\
H &  -6.364924545043 &  2.349824188415 &  0.000000000000 \\
H &   6.364924545043 & -2.349824188415 &  0.000000000000 \\
C &   0.000000000000 &  1.343476203796 &  0.000000000000 \\
C &   0.000000000000 & -1.343476203796 &  0.000000000000 \\
C &   2.346298428803 &  2.639761013996 &  0.000000000000 \\
C &  -2.346298428803 & -2.639761013996 &  0.000000000000 \\
C &  -2.346298428803 &  2.639761013996 &  0.000000000000 \\
C &   2.346298428803 & -2.639761013996 &  0.000000000000 \\
C &   4.581200368806 &  1.336477257707 &  0.000000000000 \\
C &  -4.581200368806 & -1.336477257707 &  0.000000000000 \\
C &  -4.581200368806 &  1.336477257707 &  0.000000000000 \\
C &   4.581200368806 & -1.336477257707 &  0.000000000000 \\
\end{tabular}
}
}

\vspace{5mm}

\texttt{Anthracene}

\vspace{5mm}

\footnotesize{
\texttt{
\begin{tabular}{@{}lrrr@{}}
C &    -6.890577366421 &  1.347081972134 &  0.000000000000 \\
C &    -6.890577366421 & -1.347081972134 &  0.000000000000 \\
H &    -8.678510642720 & -2.352749403648 &  0.000000000000 \\
H &    -8.678510642720 &  2.352749403648 &  0.000000000000 \\
C &    -4.672818672504 &  2.650010809649 &  0.000000000000 \\
C &    -4.672818672504 & -2.650010809649 &  0.000000000000 \\
C &    -2.302111744389 &  1.355167714500 &  0.000000000000 \\
C &    -2.302111744389 & -1.355167714500 &  0.000000000000 \\
H &    -4.664152335157 &  4.703085852735 &  0.000000000000 \\
H &    -4.664152335157 & -4.703085852735 &  0.000000000000 \\
C &     0.000000000000 &  2.638791012227 &  0.000000000000 \\
C &     0.000000000000 & -2.638791012227 &  0.000000000000 \\
C &     2.302111744389 &  1.355167714500 &  0.000000000000 \\
C &     2.302111744389 & -1.355167714500 &  0.000000000000 \\
H &     0.000000000000 &  4.693498392625 &  0.000000000000 \\
H &     0.000000000000 & -4.693498392625 &  0.000000000000 \\
C &     4.672818672504 &  2.650010809649 &  0.000000000000 \\
C &     4.672818672504 & -2.650010809649 &  0.000000000000 \\
C &     6.890577366421 &  1.347081972134 &  0.000000000000 \\
C &     6.890577366421 & -1.347081972134 &  0.000000000000 \\
H &     4.664152335157 &  4.703085852735 &  0.000000000000 \\
H &     4.664152335157 & -4.703085852735 &  0.000000000000 \\
H &     8.678510642720 &  2.352749403648 &  0.000000000000 \\
H &     8.678510642720 & -2.352749403648 &  0.000000000000 \\
\end{tabular}
}
}

\end{flushleft}
\end{table}

\begin{table}[t!]
\begin{flushleft}

\texttt{Tetracene}

\vspace{5mm}

\footnotesize{
\texttt{
\begin{tabular}{@{}lrrr@{}}
C &     2.655274746092 &   6.992957166701 &  0.000000000000 \\
C &     1.362936507814 &   4.609454612410 &  0.000000000000 \\
C &     2.645220499591 &   2.328443393116 &  0.000000000000 \\
C &     1.361090546373 &   0.000000000000 &  0.000000000000 \\
C &     1.352545447269 &   9.202892912281 &  0.000000000000 \\
H &     4.708276241841 &   6.985151156695 &  0.000000000000 \\
H &     4.699698362343 &   2.327642942569 &  0.000000000000 \\
H &     2.354402581951 &  10.992927634271 &  0.000000000000 \\
C &     2.655274746092 &  -6.992957166701 &  0.000000000000 \\
C &     1.362936507814 &  -4.609454612410 &  0.000000000000 \\
C &     2.645220499591 &  -2.328443393116 &  0.000000000000 \\
C &     1.352545447269 &  -9.202892912281 &  0.000000000000 \\
H &     4.708276241841 &  -6.985151156695 &  0.000000000000 \\
H &     4.699698362343 &  -2.327642942569 &  0.000000000000 \\
H &     2.354402581951 & -10.992927634271 &  0.000000000000 \\
C &    -2.655274746092 &   6.992957166701 &  0.000000000000 \\
C &    -1.362936507814 &   4.609454612410 &  0.000000000000 \\
C &    -2.645220499591 &   2.328443393116 &  0.000000000000 \\
C &    -1.361090546373 &   0.000000000000 &  0.000000000000 \\
C &    -1.352545447269 &   9.202892912281 &  0.000000000000 \\
H &    -4.708276241841 &   6.985151156695 &  0.000000000000 \\
H &    -4.699698362343 &   2.327642942569 &  0.000000000000 \\
H &    -2.354402581951 &  10.992927634271 &  0.000000000000 \\
C &    -2.655274746092 &  -6.992957166701 &  0.000000000000 \\
C &    -1.362936507814 &  -4.609454612410 &  0.000000000000 \\
C &    -2.645220499591 &  -2.328443393116 &  0.000000000000 \\
C &    -1.352545447269 &  -9.202892912281 &  0.000000000000 \\
H &    -4.708276241841 &  -6.985151156695 &  0.000000000000 \\
H &    -4.699698362343 &  -2.327642942569 &  0.000000000000 \\
H &    -2.354402581951 & -10.992927634271 &  0.000000000000 \\
\end{tabular}
}
}

\vspace{100mm}
\end{flushleft}
\end{table}

\newpage

\begin{table}
\begin{flushleft}

\texttt{Pentacene}

\vspace{5mm}

\footnotesize{
\texttt{
\begin{tabular}{@{}lrrr@{}}
C &     2.307152013899 & -1.366129628091 &  0.000000000000 \\
C &    -2.307152013899 & -1.366129628091 &  0.000000000000 \\
C &     2.307152013899 &  1.366129628091 &  0.000000000000 \\
C &    -2.307152013899 &  1.366129628091 &  0.000000000000 \\
C &     0.000000000000 & -2.648190127803 &  0.000000000000 \\
C &     0.000000000000 &  2.648190127803 &  0.000000000000 \\
C &     6.920529844753 &  1.367312166956 &  0.000000000000 \\
C &    -6.920529844753 &  1.367312166956 &  0.000000000000 \\
C &     6.920529844753 & -1.367312166956 &  0.000000000000 \\
C &    -6.920529844753 & -1.367312166956 &  0.000000000000 \\
C &     4.650528309668 &  2.649338730568 &  0.000000000000 \\
C &    -4.650528309668 &  2.649338730568 &  0.000000000000 \\
C &     4.650528309668 & -2.649338730568 &  0.000000000000 \\
C &    -4.650528309668 & -2.649338730568 &  0.000000000000 \\
C &    11.517833903670 &  1.355515084765 &  0.000000000000 \\
C &   -11.517833903670 &  1.355515084765 &  0.000000000000 \\
C &    11.517833903670 & -1.355515084765 &  0.000000000000 \\
C &   -11.517833903670 & -1.355515084765 &  0.000000000000 \\
C &     9.311304513754 &  2.657356094787 &  0.000000000000 \\
C &    -9.311304513754 &  2.657356094787 &  0.000000000000 \\
C &     9.311304513754 & -2.657356094787 &  0.000000000000 \\
C &    -9.311304513754 & -2.657356094787 &  0.000000000000 \\
H &     0.000000000000 & -4.702401617878 &  0.000000000000 \\
H &     0.000000000000 &  4.702401617878 &  0.000000000000 \\
H &     4.650013770338 & -4.703744534666 &  0.000000000000 \\
H &    -4.650013770338 & -4.703744534666 &  0.000000000000 \\
H &     4.650013770338 &  4.703744534666 &  0.000000000000 \\
H &    -4.650013770338 &  4.703744534666 &  0.000000000000 \\
H &    13.308757277822 & -2.355853381149 &  0.000000000000 \\
H &   -13.308757277822 & -2.355853381149 &  0.000000000000 \\
H &    13.308757277822 &  2.355853381149 &  0.000000000000 \\
H &   -13.308757277822 &  2.355853381149 &  0.000000000000 \\
H &     9.303958846038 & -4.710306035144 &  0.000000000000 \\
H &    -9.303958846038 & -4.710306035144 &  0.000000000000 \\
H &     9.303958846038 &  4.710306035144 &  0.000000000000 \\
H &    -9.303958846038 &  4.710306035144 &  0.000000000000 \\
\end{tabular}
}
}

\vspace{100mm}
\end{flushleft}
\end{table}

\begin{table}
\begin{flushleft}
\texttt{Hexacene}

\vspace{5mm}

\footnotesize{
\texttt{
\begin{tabular}{@{}lrrr@{}}
C &   -13.830141468697 &   1.357276002471 &  0.000000000000 \\ 
C &   -13.830141468697 &  -1.357276002471 &  0.000000000000 \\
H &   -15.622001573012 &  -2.355744759637 &  0.000000000000 \\
H &   -15.622001573012 &   2.355744759637 &  0.000000000000 \\
C &   -11.626906895590 &   2.660244851551 &  0.000000000000 \\
C &   -11.626906895590 &  -2.660244851551 &  0.000000000000 \\
C &    -9.232789072265 &   1.370196779459 &  0.000000000000 \\
C &    -9.232789072265 &  -1.370196779459 &  0.000000000000 \\
H &   -11.619989495045 &   4.713196046372 &  0.000000000000 \\
H &   -11.619989495045 &  -4.713196046372 &  0.000000000000 \\
C &    -6.968844729773 &   2.651354316033 &  0.000000000000 \\
C &    -6.968844729773 &  -2.651354316033 &  0.000000000000 \\
C &    -4.616598458257 &   1.369428838992 &  0.000000000000 \\
C &    -4.616598458257 &  -1.369428838992 &  0.000000000000 \\
H &    -6.968382793836 &   4.705720759567 &  0.000000000000 \\
H &    -6.968382793836 &  -4.705720759567 &  0.000000000000 \\
C &    -2.322111568177 &   2.651451011789 &  0.000000000000 \\
C &    -2.322111568177 &  -2.651451011789 &  0.000000000000 \\
C &     0.000000000000 &   1.369573946584 &  0.000000000000 \\
C &     0.000000000000 &  -1.369573946584 &  0.000000000000 \\
H &    -2.322197106783 &   4.705581795364 &  0.000000000000 \\
H &    -2.322197106783 &  -4.705581795364 &  0.000000000000 \\
C &     2.322111568177 &   2.651451011789 &  0.000000000000 \\
C &     2.322111568177 &  -2.651451011789 &  0.000000000000 \\
C &     4.616598458257 &   1.369428838992 &  0.000000000000 \\
C &     4.616598458257 &  -1.369428838992 &  0.000000000000 \\
H &     2.322197106783 &   4.705581795364 &  0.000000000000 \\
H &     2.322197106783 &  -4.705581795364 &  0.000000000000 \\
C &     6.968844729773 &   2.651354316033 &  0.000000000000 \\
C &     6.968844729773 &  -2.651354316033 &  0.000000000000 \\
C &     9.232789072265 &   1.370196779459 &  0.000000000000 \\
C &     9.232789072265 &  -1.370196779459 &  0.000000000000 \\
H &     6.968382793836 &   4.705720759567 &  0.000000000000 \\
H &     6.968382793836 &  -4.705720759567 &  0.000000000000 \\
C &    11.626906895590 &   2.660244851551 &  0.000000000000 \\
C &    11.626906895590 &  -2.660244851551 &  0.000000000000 \\
C &    13.830141468697 &   1.357276002471 &  0.000000000000 \\
C &    13.830141468697 &  -1.357276002471 &  0.000000000000 \\
H &    11.619989495045 &   4.713196046372 &  0.000000000000 \\
H &    11.619989495045 &  -4.713196046372 &  0.000000000000 \\
H &    15.622001573012 &   2.355744759637 &  0.000000000000 \\
H &    15.622001573012 &  -2.355744759637 &  0.000000000000 \\
\end{tabular}
}
}

\vspace{100mm}
\end{flushleft}
\end{table}

\begin{table}
\begin{flushleft}
\texttt{Octacene}

\vspace{5mm}

\scriptsize{
\texttt{
\begin{tabular}{@{}lrrr@{}}
C &   -18.458397509279 &   1.358953216496 &  0.000000000000 \\  
C &   -18.458397509279 &  -1.358953216496 &  0.000000000000 \\
H &   -20.250936290579 &  -2.356154956299 &  0.000000000000 \\
H &   -20.250936290579 &   2.356154956299 &  0.000000000000 \\
C &   -16.257523020858 &   2.662122767797 &  0.000000000000 \\
C &   -16.257523020858 &  -2.662122767797 &  0.000000000000 \\
C &   -13.859747750931 &   1.372937225164 &  0.000000000000 \\
C &   -13.859747750931 &  -1.372937225164 &  0.000000000000 \\
H &   -16.251051521602 &   4.715055160353 &  0.000000000000 \\
H &   -16.251051521602 &  -4.715055160353 &  0.000000000000 \\
C &   -11.601768755769 &   2.653937980769 &  0.000000000000 \\
C &   -11.601768755769 &  -2.653937980769 &  0.000000000000 \\
C &    -9.241116606364 &   1.372982540126 &  0.000000000000 \\
C &    -9.241116606364 &  -1.372982540126 &  0.000000000000 \\
H &   -11.601638174028 &   4.708270320964 &  0.000000000000 \\
H &   -11.601638174028 &  -4.708270320964 &  0.000000000000 \\
C &    -6.958245941049 &   2.654314134475 &  0.000000000000 \\
C &    -6.958245941049 &  -2.654314134475 &  0.000000000000 \\
C &    -4.620818969913 &   1.373693206486 &  0.000000000000 \\
C &    -4.620818969913 &  -1.373693206486 &  0.000000000000 \\
H &    -6.958584241451 &   4.708366406655 &  0.000000000000 \\
H &    -6.958584241451 &  -4.708366406655 &  0.000000000000 \\
C &    -2.319002610502 &   2.654853669095 &  0.000000000000 \\
C &    -2.319002610502 &  -2.654853669095 &  0.000000000000 \\
C &     0.000000000000 &   1.373925157917 &  0.000000000000 \\
C &     0.000000000000 &  -1.373925157917 &  0.000000000000 \\
H &    -2.319114781727 &   4.708851362164 &  0.000000000000 \\
H &    -2.319114781727 &  -4.708851362164 &  0.000000000000 \\
C &     2.319002610502 &   2.654853669095 &  0.000000000000 \\
C &     2.319002610502 &  -2.654853669095 &  0.000000000000 \\
C &     4.620818969913 &   1.373693206486 &  0.000000000000 \\
C &     4.620818969913 &  -1.373693206486 &  0.000000000000 \\
H &     2.319114781727 &   4.708851362164 &  0.000000000000 \\
H &     2.319114781727 &  -4.708851362164 &  0.000000000000 \\
C &     6.958245941049 &   2.654314134475 &  0.000000000000 \\
C &     6.958245941049 &  -2.654314134475 &  0.000000000000 \\
C &     9.241116606364 &   1.372982540126 &  0.000000000000 \\
C &     9.241116606364 &  -1.372982540126 &  0.000000000000 \\
H &     6.958584241451 &   4.708366406655 &  0.000000000000 \\
H &     6.958584241451 &  -4.708366406655 &  0.000000000000 \\
C &    11.601768755769 &   2.653937980769 &  0.000000000000 \\
C &    11.601768755769 &  -2.653937980769 &  0.000000000000 \\
C &    13.859747750931 &   1.372937225164 &  0.000000000000 \\
C &    13.859747750931 &  -1.372937225164 &  0.000000000000 \\
H &    11.601638174028 &   4.708270320964 &  0.000000000000 \\
H &    11.601638174028 &  -4.708270320964 &  0.000000000000 \\
C &    16.257523020858 &   2.662122767797 &  0.000000000000 \\
C &    16.257523020858 &  -2.662122767797 &  0.000000000000 \\
C &    18.458397509279 &   1.358953216496 &  0.000000000000 \\
C &    18.458397509279 &  -1.358953216496 &  0.000000000000 \\
H &    16.251051521602 &   4.715055160353 &  0.000000000000 \\
H &    16.251051521602 &  -4.715055160353 &  0.000000000000 \\
H &    20.250936290579 &   2.356154956299 &  0.000000000000 \\
H &    20.250936290579 &  -2.356154956299 &  0.000000000000 \\
\end{tabular}
}
}

\vspace{100mm}
\end{flushleft}
\end{table}

\begin{table}
\begin{flushleft}
\texttt{Coronene}

\vspace{5mm}

\footnotesize{
\texttt{
\begin{tabular}{@{}lrrr@{}}
C &     1.904310217417 &  1.904310217417 &  0.000000000000 \\  
C &     2.600934322072 & -0.696797163979 &  0.000000000000 \\
C &     0.696797163979 & -2.600934322072 &  0.000000000000 \\
C &    -1.904310217417 & -1.904310217417 &  0.000000000000 \\
C &    -2.600934322072 &  0.696797163979 &  0.000000000000 \\
C &    -0.696797163979 &  2.600934322072 &  0.000000000000 \\
C &     3.788490279285 &  3.788490279285 &  0.000000000000 \\
C &     6.369827618778 &  3.041559354466 &  0.000000000000 \\
C &     7.036551838112 &  0.551200611301 &  0.000000000000 \\
C &     5.174645873000 & -1.386580995279 &  0.000000000000 \\
C &     5.818561475307 & -3.995560940006 &  0.000000000000 \\
C &     3.995560940006 & -5.818561475307 &  0.000000000000 \\
C &     1.386580995279 & -5.174645873000 &  0.000000000000 \\
C &    -0.551200611301 & -7.036551838112 &  0.000000000000 \\
C &    -3.041559354466 & -6.369827618778 &  0.000000000000 \\
C &    -3.788490279285 & -3.788490279285 &  0.000000000000 \\
C &    -6.369827618778 & -3.041559354466 &  0.000000000000 \\
C &    -7.036551838112 & -0.551200611301 &  0.000000000000 \\
C &    -5.174645873000 &  1.386580995279 &  0.000000000000 \\
C &    -5.818561475307 &  3.995560940006 &  0.000000000000 \\
C &    -3.995560940006 &  5.818561475307 &  0.000000000000 \\
C &    -1.386580995279 &  5.174645873000 &  0.000000000000 \\
C &     0.551200611301 &  7.036551838112 &  0.000000000000 \\
C &     3.041559354466 &  6.369827618778 &  0.000000000000 \\
H &     4.499911139298 &  7.814640504403 &  0.000000000000 \\
H &     7.814640504403 &  4.499911139298 &  0.000000000000 \\
H &     9.016872803059 &  0.010291285487 &  0.000000000000 \\
H &     7.804018885964 & -4.517324239573 &  0.000000000000 \\
H &     4.517324239573 & -7.804018885964 &  0.000000000000 \\
H &    -0.010291285487 & -9.016872803059 &  0.000000000000 \\
H &    -4.499911139298 & -7.814640504403 &  0.000000000000 \\
H &    -7.814640504403 & -4.499911139298 &  0.000000000000 \\
H &    -9.016872803059 & -0.010291285487 &  0.000000000000 \\
H &    -7.804018885964 &  4.517324239573 &  0.000000000000 \\
H &    -4.517324239573 &  7.804018885964 &  0.000000000000 \\
H &     0.010291285487 &  9.016872803059 &  0.000000000000 \\
\end{tabular}
}
}

\vspace{100mm}
\end{flushleft}
\end{table}

\begin{table}
\begin{flushleft}
\texttt{BPEA}

\vspace{5mm}

\footnotesize{
\texttt{
\begin{tabular}{@{}lrrr@{}}
H &  0.000000 &  9.409086 &  0.000000 \\
H &  2.153314 &  8.159815 &  0.000000 \\
H & -2.154562 &  5.679063 &  0.000000 \\
H & -2.153314 &  8.159815 &  0.000000 \\
H &  2.154562 &  5.679063 &  0.000000 \\
C &  0.000000 &  8.323596 &  0.000000 \\
C &  0.000000 &  5.502856 &  0.000000 \\
C &  1.211510 &  7.619661 &  0.000000 \\
C & -1.216455 &  6.224237 &  0.000000 \\
C & -1.211510 &  7.619661 &  0.000000 \\
C &  1.216455 &  6.224237 &  0.000000 \\
C &  0.000000 &  2.854242 &  0.000000 \\
C &  0.000000 &  4.077836 &  0.000000 \\
C & -3.680574 &  0.711463 &  0.000000 \\
C & -3.680574 & -0.711463 &  0.000000 \\
H & -4.623556 & -1.249145 &  0.000000 \\
H & -4.623556 &  1.249145 &  0.000000 \\
C & -2.492116 &  1.403139 &  0.000000 \\
C & -2.492116 & -1.403139 &  0.000000 \\
C & -1.235157 &  0.720506 &  0.000000 \\
C & -1.235157 & -0.720506 &  0.000000 \\
H & -2.486136 &  2.487554 &  0.000000 \\
H & -2.486136 & -2.487554 &  0.000000 \\
C &  0.000000 &  1.434153 &  0.000000 \\
C &  0.000000 & -1.434153 &  0.000000 \\
C &  1.235157 &  0.720506 &  0.000000 \\
C &  1.235157 & -0.720506 &  0.000000 \\
C &  2.492116 &  1.403139 &  0.000000 \\
C &  2.492116 & -1.403139 &  0.000000 \\
C &  3.680574 &  0.711463 &  0.000000 \\
C &  3.680574 & -0.711463 &  0.000000 \\
H &  2.486136 &  2.487554 &  0.000000 \\
H &  2.486136 & -2.487554 &  0.000000 \\
H &  4.623556 &  1.249145 &  0.000000 \\
H &  4.623556 & -1.249145 &  0.000000 \\
C &  0.000000 & -2.854242 &  0.000000 \\
C &  0.000000 & -4.077836 &  0.000000 \\
H &  0.000000 & -9.409086 &  0.000000 \\
H &  2.153314 & -8.159815 &  0.000000 \\
H & -2.154562 & -5.679063 &  0.000000 \\
H & -2.153314 & -8.159815 &  0.000000 \\
H &  2.154562 & -5.679063 &  0.000000 \\
C &  0.000000 & -8.323596 &  0.000000 \\
C &  0.000000 & -5.502856 &  0.000000 \\
C &  1.211510 & -7.619661 &  0.000000 \\
C & -1.216455 & -6.224237 &  0.000000 \\
C & -1.211510 & -7.619661 &  0.000000 \\
C &  1.216455 & -6.224237 &  0.000000 \\
\end{tabular}
}
}

\vspace{100mm}
\end{flushleft}
\end{table}

\begin{table}
\begin{flushleft}

{\large \textbf{Trans-polyacetylene (TPA) molecules}}

\vspace{5mm}

\texttt{C$_{8}$H$_{10}$}

\vspace{5mm}

\footnotesize{
\texttt{
\begin{tabular}{@{}lrrr@{}}
C &  0.72500000000 &  0.00000000000 &  0.00000000000 \\
C & -0.72500000000 &  0.00000000000 &  0.00000000000 \\
C &  1.39500000000 &  1.16047404107 &  0.00000000000 \\
C & -1.39500000000 & -1.16047404107 &  0.00000000000 \\
H &  1.26500000000 & -0.93530743609 &  0.00000000000 \\
H & -1.26500000000 &  0.93530743609 &  0.00000000000 \\
C &  2.84500000000 &  1.16047404107 &  0.00000000000 \\
C & -2.84500000000 & -1.16047404107 &  0.00000000000 \\
H &  0.85500000000 &  2.09578147716 &  0.00000000000 \\
H & -0.85500000000 & -2.09578147716 &  0.00000000000 \\
C &  3.51500000000 &  2.32094808214 &  0.00000000000 \\
C & -3.51500000000 & -2.32094808214 &  0.00000000000 \\
H &  3.38500000000 &  0.22516660498 &  0.00000000000 \\
H & -3.38500000000 & -0.22516660498 &  0.00000000000 \\
H &  4.59500000000 &  2.32094808214 &  0.00000000000 \\
H & -4.59500000000 & -2.32094808214 &  0.00000000000 \\
H &  2.97500000000 &  3.25625551823 &  0.00000000000 \\
H & -2.97500000000 & -3.25625551823 &  0.00000000000 \\
\end{tabular}
}
}

\vspace{10mm}

\texttt{C$_{12}$H$_{14}$}

\vspace{5mm}

\footnotesize{
\texttt{
\begin{tabular}{@{}lrrr@{}}
C &  0.72500000000 &  0.00000000000 &  0.00000000000 \\
C & -0.72500000000 &  0.00000000000 &  0.00000000000 \\
C &  1.39500000000 &  1.16047404107 &  0.00000000000 \\
C & -1.39500000000 & -1.16047404107 &  0.00000000000 \\
H &  1.26500000000 & -0.93530743609 &  0.00000000000 \\
H & -1.26500000000 &  0.93530743609 &  0.00000000000 \\
C &  2.84500000000 &  1.16047404107 &  0.00000000000 \\
C & -2.84500000000 & -1.16047404107 &  0.00000000000 \\
H &  0.85500000000 &  2.09578147716 &  0.00000000000 \\
H & -0.85500000000 & -2.09578147716 &  0.00000000000 \\
C &  3.51500000000 &  2.32094808214 &  0.00000000000 \\
C & -3.51500000000 & -2.32094808214 &  0.00000000000 \\
H &  3.38500000000 &  0.22516660498 &  0.00000000000 \\
H & -3.38500000000 & -0.22516660498 &  0.00000000000 \\
C &  4.96500000000 &  2.32094808214 &  0.00000000000 \\
C & -4.96500000000 & -2.32094808214 &  0.00000000000 \\
H &  2.97500000000 &  3.25625551823 &  0.00000000000 \\
H & -2.97500000000 & -3.25625551823 &  0.00000000000 \\
C &  5.63500000000 &  3.48142212321 &  0.00000000000 \\
C & -5.63500000000 & -3.48142212321 &  0.00000000000 \\
H &  5.50500000000 &  1.38564064606 &  0.00000000000 \\
H & -5.50500000000 & -1.38564064606 &  0.00000000000 \\
H &  6.71500000000 &  3.48142212321 &  0.00000000000 \\
H & -6.71500000000 & -3.48142212321 &  0.00000000000 \\
H &  5.09500000000 &  4.41672955930 &  0.00000000000 \\
H & -5.09500000000 & -4.41672955930 &  0.00000000000 \\
\end{tabular}
}
}

\end{flushleft}
\end{table}

\begin{table}[t!]
\begin{flushleft}

\vspace{5mm}

\texttt{C$_{16}$H$_{18}$}

\vspace{5mm}

\footnotesize{
\texttt{
\begin{tabular}{@{}lrrr@{}}
C &  0.72500000000 &  0.00000000000 &  0.00000000000 \\
C & -0.72500000000 &  0.00000000000 &  0.00000000000 \\
C &  1.39500000000 &  1.16047404107 &  0.00000000000 \\
C & -1.39500000000 & -1.16047404107 &  0.00000000000 \\
H &  1.26500000000 & -0.93530743609 &  0.00000000000 \\
H & -1.26500000000 &  0.93530743609 &  0.00000000000 \\
C &  2.84500000000 &  1.16047404107 &  0.00000000000 \\
C & -2.84500000000 & -1.16047404107 &  0.00000000000 \\
H &  0.85500000000 &  2.09578147716 &  0.00000000000 \\
H & -0.85500000000 & -2.09578147716 &  0.00000000000 \\
C &  3.51500000000 &  2.32094808214 &  0.00000000000 \\
C & -3.51500000000 & -2.32094808214 &  0.00000000000 \\
H &  3.38500000000 &  0.22516660498 &  0.00000000000 \\
H & -3.38500000000 & -0.22516660498 &  0.00000000000 \\
C &  4.96500000000 &  2.32094808214 &  0.00000000000 \\
C & -4.96500000000 & -2.32094808214 &  0.00000000000 \\
H &  2.97500000000 &  3.25625551823 &  0.00000000000 \\
H & -2.97500000000 & -3.25625551823 &  0.00000000000 \\
C &  5.63500000000 &  3.48142212321 &  0.00000000000 \\
C & -5.63500000000 & -3.48142212321 &  0.00000000000 \\
H &  5.50500000000 &  1.38564064606 &  0.00000000000 \\
H & -5.50500000000 & -1.38564064606 &  0.00000000000 \\
C &  7.08500000000 &  3.48142212321 &  0.00000000000 \\
C & -7.08500000000 & -3.48142212321 &  0.00000000000 \\
H &  5.09500000000 &  4.41672955930 &  0.00000000000 \\
H & -5.09500000000 & -4.41672955930 &  0.00000000000 \\
C &  7.75500000000 &  4.64189616428 &  0.00000000000 \\
C & -7.75500000000 & -4.64189616428 &  0.00000000000 \\
H &  7.62500000000 &  2.54611468713 &  0.00000000000 \\
H & -7.62500000000 & -2.54611468713 &  0.00000000000 \\
H &  8.83500000000 &  4.64189616428 &  0.00000000000 \\
H & -8.83500000000 & -4.64189616428 &  0.00000000000 \\
H &  7.21500000000 &  5.57720360037 &  0.00000000000 \\
H & -7.21500000000 & -5.57720360037 &  0.00000000000 \\
\end{tabular}
}
}

\vspace{100mm}
\end{flushleft}
\end{table}

\begin{table}[t!]
\begin{flushleft}

\texttt{C$_{20}$H$_{22}$}

\vspace{5mm}

\footnotesize{
\texttt{
\begin{tabular}{@{}lrrr@{}}
C &  0.72500000000 &  0.00000000000 &  0.00000000000 \\
C & -0.72500000000 &  0.00000000000 &  0.00000000000 \\
C &  1.39500000000 &  1.16047404107 &  0.00000000000 \\
C & -1.39500000000 & -1.16047404107 &  0.00000000000 \\
H &  1.26500000000 & -0.93530743609 &  0.00000000000 \\
H & -1.26500000000 &  0.93530743609 &  0.00000000000 \\
C &  2.84500000000 &  1.16047404107 &  0.00000000000 \\
C & -2.84500000000 & -1.16047404107 &  0.00000000000 \\
H &  0.85500000000 &  2.09578147716 &  0.00000000000 \\
H & -0.85500000000 & -2.09578147716 &  0.00000000000 \\
C &  3.51500000000 &  2.32094808214 &  0.00000000000 \\
C & -3.51500000000 & -2.32094808214 &  0.00000000000 \\
H &  3.38500000000 &  0.22516660498 &  0.00000000000 \\
H & -3.38500000000 & -0.22516660498 &  0.00000000000 \\
C &  4.96500000000 &  2.32094808214 &  0.00000000000 \\
C & -4.96500000000 & -2.32094808214 &  0.00000000000 \\
H &  2.97500000000 &  3.25625551823 &  0.00000000000 \\
H & -2.97500000000 & -3.25625551823 &  0.00000000000 \\
C &  5.63500000000 &  3.48142212321 &  0.00000000000 \\
C & -5.63500000000 & -3.48142212321 &  0.00000000000 \\
H &  5.50500000000 &  1.38564064606 &  0.00000000000 \\
H & -5.50500000000 & -1.38564064606 &  0.00000000000 \\
C &  7.08500000000 &  3.48142212321 &  0.00000000000 \\
C & -7.08500000000 & -3.48142212321 &  0.00000000000 \\
H &  5.09500000000 &  4.41672955930 &  0.00000000000 \\
H & -5.09500000000 & -4.41672955930 &  0.00000000000 \\
C &  7.75500000000 &  4.64189616428 &  0.00000000000 \\
C & -7.75500000000 & -4.64189616428 &  0.00000000000 \\
H &  7.62500000000 &  2.54611468713 &  0.00000000000 \\
H & -7.62500000000 & -2.54611468713 &  0.00000000000 \\
C &  9.20500000000 &  4.64189616428 &  0.00000000000 \\
C & -9.20500000000 & -4.64189616428 &  0.00000000000 \\
H &  7.21500000000 &  5.57720360037 &  0.00000000000 \\
H & -7.21500000000 & -5.57720360037 &  0.00000000000 \\
C &  9.87500000000 &  5.80237020536 &  0.00000000000 \\
C & -9.87500000000 & -5.80237020536 &  0.00000000000 \\
H &  9.74500000000 &  3.70658872820 &  0.00000000000 \\
H & -9.74500000000 & -3.70658872820 &  0.00000000000 \\
H &  10.9550000000 &  5.80237020536 &  0.00000000000 \\
H & -10.9550000000 & -5.80237020536 &  0.00000000000 \\
H &  9.33500000000 &  6.73767764144 &  0.00000000000 \\
H & -9.33500000000 & -6.73767764144 &  0.00000000000 \\
\end{tabular}
}
}

\vspace{100mm}
\end{flushleft}
\end{table}

\begin{table}[t!]
\begin{flushleft}

\texttt{C$_{24}$H$_{26}$}

\vspace{5mm}

\footnotesize{
\texttt{
\begin{tabular}{@{}lrrr@{}}
C &  0.72500000000 &  0.00000000000 &  0.00000000000 \\
C & -0.72500000000 &  0.00000000000 &  0.00000000000 \\
C &  1.39500000000 &  1.16047404107 &  0.00000000000 \\
C & -1.39500000000 & -1.16047404107 &  0.00000000000 \\
H &  1.26500000000 & -0.93530743609 &  0.00000000000 \\
H & -1.26500000000 &  0.93530743609 &  0.00000000000 \\
C &  2.84500000000 &  1.16047404107 &  0.00000000000 \\
C & -2.84500000000 & -1.16047404107 &  0.00000000000 \\
H &  0.85500000000 &  2.09578147716 &  0.00000000000 \\
H & -0.85500000000 & -2.09578147716 &  0.00000000000 \\
C &  3.51500000000 &  2.32094808214 &  0.00000000000 \\
C & -3.51500000000 & -2.32094808214 &  0.00000000000 \\
H &  3.38500000000 &  0.22516660498 &  0.00000000000 \\
H & -3.38500000000 & -0.22516660498 &  0.00000000000 \\
C &  4.96500000000 &  2.32094808214 &  0.00000000000 \\
C & -4.96500000000 & -2.32094808214 &  0.00000000000 \\
H &  2.97500000000 &  3.25625551823 &  0.00000000000 \\
H & -2.97500000000 & -3.25625551823 &  0.00000000000 \\
C &  5.63500000000 &  3.48142212321 &  0.00000000000 \\
C & -5.63500000000 & -3.48142212321 &  0.00000000000 \\
H &  5.50500000000 &  1.38564064606 &  0.00000000000 \\
H & -5.50500000000 & -1.38564064606 &  0.00000000000 \\
C &  7.08500000000 &  3.48142212321 &  0.00000000000 \\
C & -7.08500000000 & -3.48142212321 &  0.00000000000 \\
H &  5.09500000000 &  4.41672955930 &  0.00000000000 \\
H & -5.09500000000 & -4.41672955930 &  0.00000000000 \\
C &  7.75500000000 &  4.64189616428 &  0.00000000000 \\
C & -7.75500000000 & -4.64189616428 &  0.00000000000 \\
H &  7.62500000000 &  2.54611468713 &  0.00000000000 \\
H & -7.62500000000 & -2.54611468713 &  0.00000000000 \\
C &  9.20500000000 &  4.64189616428 &  0.00000000000 \\
C & -9.20500000000 & -4.64189616428 &  0.00000000000 \\
H &  7.21500000000 &  5.57720360037 &  0.00000000000 \\
H & -7.21500000000 & -5.57720360037 &  0.00000000000 \\
C &  9.87500000000 &  5.80237020536 &  0.00000000000 \\
C & -9.87500000000 & -5.80237020536 &  0.00000000000 \\
H &  9.74500000000 &  3.70658872820 &  0.00000000000 \\
H & -9.74500000000 & -3.70658872820 &  0.00000000000 \\
C &  11.3250000000 &  5.80237020536 &  0.00000000000 \\
C & -11.3250000000 & -5.80237020536 &  0.00000000000 \\
H &  9.33500000000 &  6.73767764144 &  0.00000000000 \\
H & -9.33500000000 & -6.73767764144 &  0.00000000000 \\
C &  11.9950000000 &  6.96284424643 &  0.00000000000 \\
C & -11.9950000000 & -6.96284424643 &  0.00000000000 \\
H &  11.8650000000 &  4.86706276927 &  0.00000000000 \\
H & -11.8650000000 & -4.86706276927 &  0.00000000000 \\
H &  13.0750000000 &  6.96284424643 &  0.00000000000 \\
H & -13.0750000000 & -6.96284424643 &  0.00000000000 \\
H &  11.4550000000 &  7.89815168251 &  0.00000000000 \\
H & -11.4550000000 & -7.89815168251 &  0.00000000000 \\
\end{tabular}
}
}

\vspace{100mm}
\end{flushleft}
\end{table}

\begin{table}[t!]
\begin{flushleft}

\texttt{C$_{28}$H$_{30}$}

\vspace{5mm}

\scriptsize{
\texttt{
\begin{tabular}{@{}lrrr@{}}
C &  0.72500000000 &  0.00000000000 &  0.00000000000 \\
C & -0.72500000000 &  0.00000000000 &  0.00000000000 \\
C &  1.39500000000 &  1.16047404107 &  0.00000000000 \\
C & -1.39500000000 & -1.16047404107 &  0.00000000000 \\
H &  1.26500000000 & -0.93530743609 &  0.00000000000 \\
H & -1.26500000000 &  0.93530743609 &  0.00000000000 \\
C &  2.84500000000 &  1.16047404107 &  0.00000000000 \\
C & -2.84500000000 & -1.16047404107 &  0.00000000000 \\
H &  0.85500000000 &  2.09578147716 &  0.00000000000 \\
H & -0.85500000000 & -2.09578147716 &  0.00000000000 \\
C &  3.51500000000 &  2.32094808214 &  0.00000000000 \\
C & -3.51500000000 & -2.32094808214 &  0.00000000000 \\
H &  3.38500000000 &  0.22516660498 &  0.00000000000 \\
H & -3.38500000000 & -0.22516660498 &  0.00000000000 \\
C &  4.96500000000 &  2.32094808214 &  0.00000000000 \\
C & -4.96500000000 & -2.32094808214 &  0.00000000000 \\
H &  2.97500000000 &  3.25625551823 &  0.00000000000 \\
H & -2.97500000000 & -3.25625551823 &  0.00000000000 \\
C &  5.63500000000 &  3.48142212321 &  0.00000000000 \\
C & -5.63500000000 & -3.48142212321 &  0.00000000000 \\
H &  5.50500000000 &  1.38564064606 &  0.00000000000 \\
H & -5.50500000000 & -1.38564064606 &  0.00000000000 \\
C &  7.08500000000 &  3.48142212321 &  0.00000000000 \\
C & -7.08500000000 & -3.48142212321 &  0.00000000000 \\
H &  5.09500000000 &  4.41672955930 &  0.00000000000 \\
H & -5.09500000000 & -4.41672955930 &  0.00000000000 \\
C &  7.75500000000 &  4.64189616428 &  0.00000000000 \\
C & -7.75500000000 & -4.64189616428 &  0.00000000000 \\
H &  7.62500000000 &  2.54611468713 &  0.00000000000 \\
H & -7.62500000000 & -2.54611468713 &  0.00000000000 \\
C &  9.20500000000 &  4.64189616428 &  0.00000000000 \\
C & -9.20500000000 & -4.64189616428 &  0.00000000000 \\
H &  7.21500000000 &  5.57720360037 &  0.00000000000 \\
H & -7.21500000000 & -5.57720360037 &  0.00000000000 \\
C &  9.87500000000 &  5.80237020536 &  0.00000000000 \\
C & -9.87500000000 & -5.80237020536 &  0.00000000000 \\
H &  9.74500000000 &  3.70658872820 &  0.00000000000 \\
H & -9.74500000000 & -3.70658872820 &  0.00000000000 \\
C &  11.3250000000 &  5.80237020536 &  0.00000000000 \\
C & -11.3250000000 & -5.80237020536 &  0.00000000000 \\
H &  9.33500000000 &  6.73767764144 &  0.00000000000 \\
H & -9.33500000000 & -6.73767764144 &  0.00000000000 \\
C &  11.9950000000 &  6.96284424643 &  0.00000000000 \\
C & -11.9950000000 & -6.96284424643 &  0.00000000000 \\
H &  11.8650000000 &  4.86706276927 &  0.00000000000 \\
H & -11.8650000000 & -4.86706276927 &  0.00000000000 \\
C &  13.4450000000 &  6.96284424643 &  0.00000000000 \\
C & -13.4450000000 & -6.96284424643 &  0.00000000000 \\
H &  11.4550000000 &  7.89815168251 &  0.00000000000 \\
H & -11.4550000000 & -7.89815168251 &  0.00000000000 \\
C &  14.1150000000 &  8.12331828750 &  0.00000000000 \\
C & -14.1150000000 & -8.12331828750 &  0.00000000000 \\
H &  13.9850000000 &  6.02753681034 &  0.00000000000 \\
H & -13.9850000000 & -6.02753681034 &  0.00000000000 \\
H &  15.1950000000 &  8.12331828750 &  0.00000000000 \\
H & -15.1950000000 & -8.12331828750 &  0.00000000000 \\
H &  13.5750000000 &  9.05862572359 &  0.00000000000 \\
H & -13.5750000000 & -9.05862572359 &  0.00000000000 \\
\end{tabular}
}
}

\vspace{100mm}
\end{flushleft}
\end{table}

\begin{table}[t!]
\begin{flushleft}

{\large \textbf{Other systems}}

\vspace{5mm}

\texttt{Ferrocene}

\vspace{5mm}

\footnotesize{
\texttt{
\begin{tabular}{@{}lrrr@{}}
Fe &    0.000000 &   0.000000 &   0.000000 \\
C  &   -0.713500 &  -0.982049 &  -1.648000 \\ 
C  &    0.713500 &  -0.982049 &  -1.648000 \\ 
C  &    1.154467 &   0.375109 &  -1.648000 \\ 
C  &    0.000000 &   1.213879 &  -1.648000 \\ 
C  &   -1.154467 &   0.375109 &  -1.648000 \\ 
H  &   -1.347694 &  -1.854942 &  -1.638208 \\ 
H  &    1.347694 &  -1.854942 &  -1.638208 \\ 
H  &    2.180615 &   0.708525 &  -1.638208 \\ 
H  &    0.000000 &   2.292835 &  -1.638208 \\ 
H  &   -2.180615 &   0.708525 &  -1.638208 \\ 
C  &   -0.713500 &  -0.982049 &   1.648000 \\ 
C  &   -1.154467 &   0.375109 &   1.648000 \\ 
C  &   -0.000000 &   1.213879 &   1.648000 \\ 
C  &    1.154467 &   0.375109 &   1.648000 \\ 
C  &    0.713500 &  -0.982049 &   1.648000 \\ 
H  &   -1.347694 &  -1.854942 &   1.638208 \\ 
H  &   -2.180615 &   0.708525 &   1.638208 \\ 
H  &    0.000000 &   2.292835 &   1.638208 \\ 
H  &    2.180615 &   0.708525 &   1.638208 \\ 
H  &    1.347694 &  -1.854942 &   1.638208 \\
\end{tabular}
}
}

\vspace{200mm}
\end{flushleft}
\end{table}

\newpage

\begin{table}[t!]
\begin{flushleft}

\texttt{Fe(II)-Porphyrin}

\vspace{5mm}

\footnotesize{
\texttt{
\begin{tabular}{@{}lrrr@{}}
Fe & 0.0000 & 0.0000 & 0.0000 \\
N & 1.9764 & 0.0000 & 0.0000 \\
N & 0.0000 & 1.9884 & 0.0000 \\
N & -1.9764 & 0.0000 & 0.0000 \\
N & 0.0000 & -1.9884 & 0.0000 \\
C & 2.8182 & -1.0903 & 0.0000 \\
C & 2.8182 & 1.0903 & 0.0000 \\
C & 1.0918 & 2.8249 & 0.0000 \\
C & -1.0918 & 2.8249 & 0.0000 \\
C & -2.8182 & 1.0903 & 0.0000 \\
C & -2.8182 & -1.0903 & 0.0000 \\
C & -1.0918 & -2.8249 & 0.0000 \\
C & 1.0918 & -2.8249 & 0.0000 \\
C & 4.1961 & -0.6773 & 0.0000 \\
C & 4.1961 & 0.6773 & 0.0000 \\
C & 0.6825 & 4.1912 & 0.0000 \\
C & -0.6825 & 4.1912 & 0.0000 \\
C & -4.1961 & 0.6773 & 0.0000 \\
C & -4.1961 & -0.6773 & 0.0000 \\
C & -0.6825 & -4.1912 & 0.0000 \\
C & 0.6825 & -4.1912 & 0.0000 \\
H & 5.0441 & -1.3538 & 0.0000 \\
H & 5.0441 & 1.3538 & 0.0000 \\
H & 1.3558 & 5.0416 & 0.0000 \\
H & -1.3558 & 5.0416 & 0.0000 \\
H & -5.0441 & 1.3538 & 0.0000 \\
H & -5.0441 & -1.3538 & 0.0000 \\
H & -1.3558 & -5.0416 & 0.0000 \\
H & 1.3558 & -5.0416 & 0.0000 \\
C & 2.4150 & 2.4083 & 0.0000 \\
C & -2.4150 & 2.4083 & 0.0000 \\
C & -2.4150 & -2.4083 & 0.0000 \\
C & 2.4150 & -2.4083 & 0.0000 \\
H & 3.1855 & 3.1752 & 0.0000 \\
H & -3.1855 & 3.1752 & 0.0000 \\
H & -3.1855 & -3.1752 & 0.0000 \\
H & 3.1855 & -3.1752 & 0.0000 \\
\end{tabular}
}
}

\vspace{100mm}
\end{flushleft}
\end{table}

\clearpage

\bibliography{supp_material}